\documentclass[aps,prx,reprint,superscriptaddress]{revtex4-2}

\usepackage{graphicx}
\usepackage{enumerate}
\usepackage{soul}
\usepackage{amsmath}
\usepackage{amssymb}
\usepackage{gensymb}
\usepackage{appendix}
\usepackage{hyperref}

\hypersetup{
    colorlinks=true,
    allcolors=blue,
}

\begin{document}

\title{Visualising energy landscapes through manifold learning}

\author{Ben W.\ B.\ Shires}
\email[]{bs511@cam.ac.uk}
\affiliation{Department of Materials Science \& Metallurgy, University of Cambridge, 27 Charles Babbage Road, Cambridge CB3~0FS, United Kingdom}

\author{Chris J.\ Pickard}
\email[]{cjp20@cam.ac.uk}
\affiliation{Department of Materials Science \& Metallurgy, University of Cambridge, 27 Charles Babbage Road, Cambridge CB3~0FS, United Kingdom}
\affiliation{Advanced Institute for Materials Research, Tohoku University 2-1-1 Katahira, Aoba, Sendai, 980-8577, Japan}

\date{\today}

\begin{abstract}
Energy landscapes provide a conceptual framework for structure prediction, and a detailed understanding of their topological features is necessary to develop efficient methods for their exploration. The ability to visualise these surfaces is essential, but the high dimensionality of the corresponding configuration spaces makes this difficult. Here we present Stochastic Hyperspace Embedding and Projection (SHEAP), a method for energy landscape visualisation inspired by state-of-the-art algorithms for dimensionality reduction through manifold learning, such as t-SNE and UMAP. The performance of SHEAP is demonstrated through its application to the energy landscapes of Lennard-Jones clusters, solid-state carbon, and the quaternary system C+H+N+O. It produces meaningful and interpretable low-dimensional representations of these landscapes, reproducing well known topological features such as funnels, and providing fresh insight into their layouts. In particular, an intrinsic low dimensionality in the distribution of local minima across configuration space is revealed.
\end{abstract}


\maketitle

\section{Introduction}
\label{introduction}

Potential (and free) energy surfaces (PESs) provide a conceptual and mathematical framework for describing the structures, kinetics and thermodynamics of systems of interacting atoms \cite{born1927zur,wales2003energy,oganov2009quantify,wales2012decoding}. The exploration of these surfaces forms the basis for the field of (computational) structure prediction, where the central task is the location of low-lying minima \cite{oganov2019structure}. In order to develop efficient methods for their exploration, the ability to visualise these landscapes is crucial, enhancing our understanding of their complex topologies and providing insight into how we can improve techniques for structure prediction and global optimisation. However, the potential energy of a system of atoms varies with the relative positions of all of its nuclei, meaning that for all but the simplest of systems the true surface cannot be depicted in the three spatial dimensions that we can easily visualise.

There are many successful approaches to structure prediction, from minima/basin-hopping \cite{li1987monte,wales1997global,amsler2010crystal},  particle swarm \cite{wang2010crystal,wang2012calypso}, and evolutionary algorithms \cite{oganov2006crystal,glass2006uspex}, to ab initio random structure searching (AIRSS) \cite{pickard2006high,pickard2011ab}. These methods generate a diverse range of structures for any given system, and AIRSS, in particular, emphasises a broad sampling of the energy landscape. This approach has been successful in identifying novel and interesting phases of crystalline systems, particularly at high pressures \cite{pickard2007structure,pickard2008highly,pickard2010aluminium}. Such a random search for local minima produces meaningful samples of configuration space, relating directly to the underlying distribution of the basins of attraction. However, these samples are generated in huge quantities (around $100{,}000$ for modern application), and the resulting data-sets can be hard to fully comprehend, making it difficult to determine how a given search might be improved. A means to extract and depict information contained in these large data-sets of local minima, such that insight into the landscape being searched can be obtained, would be valuable.

The desire to visualise high-dimensional objects is not restricted to energy landscapes. A similar challenge is encountered for loss functions defining other optimisation tasks, such as statistical regression schemes \cite{kleinbaum1988applied}, and machine learning algorithms for classification, clustering, and neural networks \cite{jain1999data,aggarwal2014data,nielsen2015neural,chitturi2020perspective}. For any such cost function surface, it would be revealing if an approximate low-dimensional representation could be produced which preserves as faithfully as possible the key features of the landscape, such as the location of stationary points. For the PESs sampled by methods for structure prediction, the important features include the relative layout of the local minima, the connectivity of the corresponding basins of attraction, the relative volumes of the basins, and the presence of ``super-basins'', or funnels \cite{wales2003energy,stillinger1984packing,doye2005characterizing,massen2007power,bryngelson1995funnels,becker1997topology,oganov2009quantify}. Here we propose a method for energy landscape visualisation which captures these features.

PES visualisation can be viewed as a data science task in which we would like to project a high-dimensional source data-set (consisting of vectors describing structures in the configuration space of interest) into one, two or three Euclidean dimensions. We base our approach on state-of-the-art algorithms for dimensionality reduction for visualisation, such as t-Distributed Stochastic Neighbour Embedding (t-SNE) \cite{maaten2008visualizing} and Uniform Manifold Approximation and Projection (UMAP) \cite{mcinnes2018umap1,mcinnes2018umap2}, drawing inspiration from their increasing application to data-sets arising in physics \cite{fosel2018reinforcement,li2019manifold,spiwok2020time}, materials science \cite{olsthoorn2019band,royse2019emergence}, and cell biology \cite{ghandi2019next,dutertre2019single,hemmers20192,becht2019dimensionality}. We call this approach SHEAP - Stochastic Hyperspace Embedding And Projection.

There are a number of existing tools which allow the depiction of certain features of high-dimensional PESs in lower dimensions, including (metric) disconnectivity graphs \cite{czerminski1990reaction,becker1997topology,wales2004energy,rylance2006topographical,smeeton2014visualizing}, and sketch-maps \cite{ceriotti2011simplifying,tribello2012using,ceriotti2013demonstrating}. These approaches do not focus on faithfully preserving the spatial layout of minima on the true PES, but rather on understanding the system dynamics.

Disconnectivity graphs display the connectivity of local minima in terms of minimum energy pathways between them, illustrating which structures are likely to be accessible and interchangeable at a given temperature \cite{czerminski1990reaction,becker1997topology,wales2004energy}. They do not indicate the proximity of minima in the configuration space according to some defined distance metric, which basins are actually adjacent on the PES, or give a clear indication of the relative basin volumes \cite{rylance2006topographical}.

Metric disconnectivity graphs retain some structural information about the minima by arranging them along one or two metric axes \cite{rylance2006topographical,smeeton2014visualizing}. However, since they use one of the two or three plot dimensions to represent the energies of the structures depicted, there is one fewer dimension which can be used to represent the relative positions of the minima on the surface. Furthermore, techniques such as t-SNE and UMAP have been demonstrated to outperform the dimensionality reduction methods considered for metric disconnectivity graphs in the context of visualisation \cite{maaten2008visualizing,mcinnes2018umap1}.

Sketch-map follows a similar framework to the state-of-the-art methods that we build on here to produce SHEAP (see Section \ref{dim_red}). The algorithmic choices made within this framework deal with issues specific to the high-dimensional data produced in molecular dynamics trajectories, such as poor sampling at the transition states, and the high-dimensional nature of thermal fluctuations. Sketch-map ensures that data-points from the same basin are close, and separates clusters of structures from different basins \cite{ceriotti2011simplifying,ceriotti2013demonstrating}. However, it puts less emphasis on accurately reproducing the distances between the structures from different basins. Thus, the approach is less suited to dealing with other forms of structural data, such as the minima produced from a structure search, for which there are no thermal fluctuations (although sketch-map has been applied to this type of data with success \cite{engel2018mapping}).

In the next section we introduce dimensionality reduction in the context of visualisation, focusing on manifold learning approaches. This is followed by an outline of our method, SHEAP, designed for PES visualisation. We explore the performance of SHEAP by applying it first to model systems of atoms interacting through a Lennard-Jones (LJ) pair potential, then to structures of solid-state carbon, as well as the quaternary system C+H+N+O. We show that SHEAP yields meaningful low-dimensional representations of these landscapes, for example reproducing well known topological features such as funnels, using only the structural information contained in minima data obtained from random structure searches. Furthermore, and importantly, we reveal an intrinsic low dimensionality to the distribution of minima across these surfaces.

\section{Dimensionality reduction by manifold learning}\label{dim_red}

\begin{figure}
    \centering
    \includegraphics[width=\linewidth]{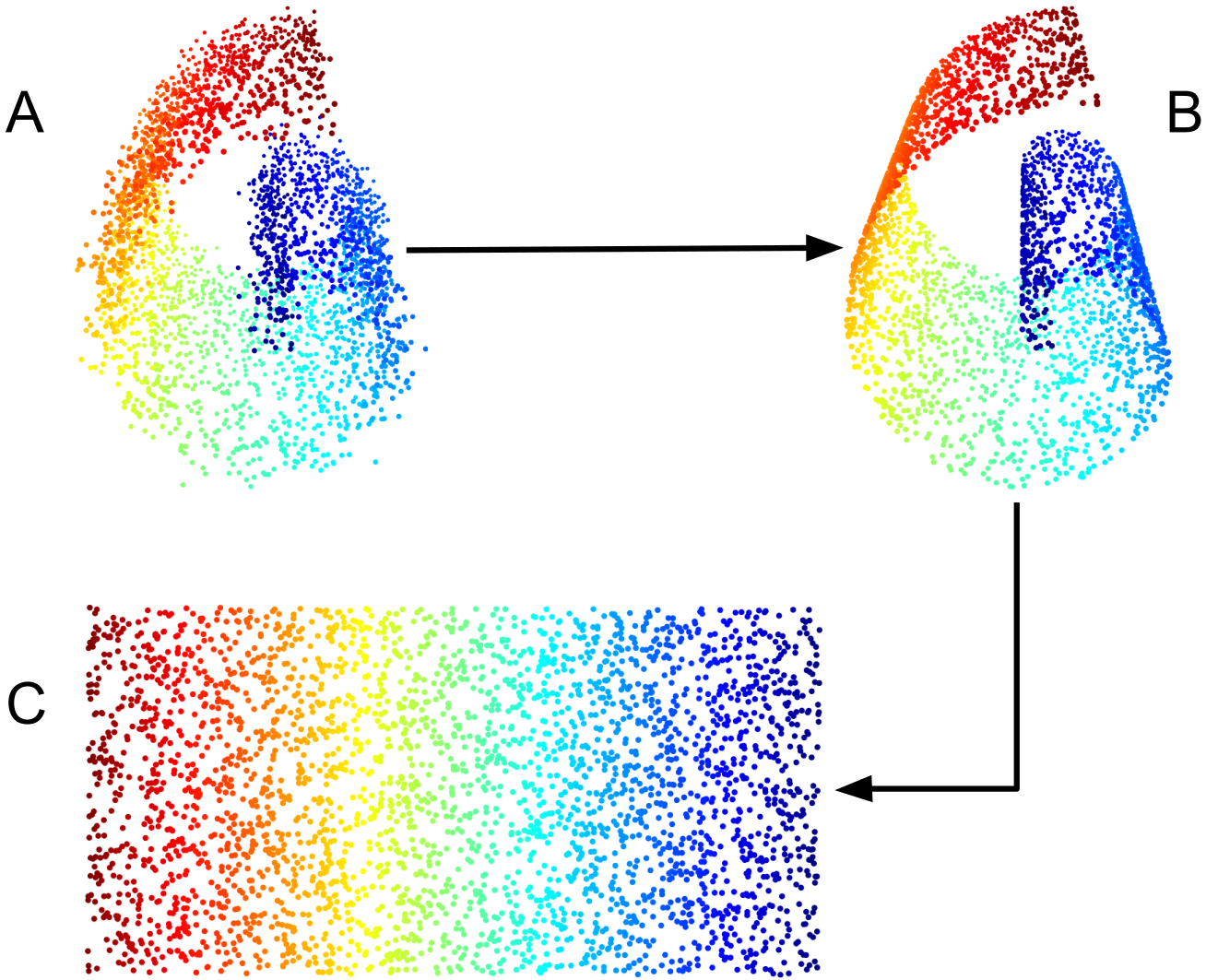}
    \caption{Illustrates data dimensionality reduction by manifold learning. (A) shows a 3-dimensional data-set that is close to lying on a 2-dimensional surface which is embedded within $\mathbb{R}^3$. (B) shows the projection of the source data onto this embedded manifold, which is folded into a swiss roll in the original 3D space. (C) shows the distribution of the data across the embedded manifold, providing a 2D visual representation of the source data.}
    \label{fig:dim_red}
\end{figure}

Dimensionality reduction refers to the mapping of a high-dimensional data-set into a space of lower dimensionality \cite{van2009dimensionality}. Here we refer to a source data-set $X = \{ \mathbf{x}_1 , \mathbf{x}_2, \dots , \mathbf{x}_N \} \subset \mathbb{R}^D$ and a low-dimensional representation $Y = \{ \mathbf{y}_1 , \mathbf{y}_2, \dots , \mathbf{y}_N \} \subset \mathbb{R}^s$, where $N$ is the number of data points, and $D$ and $s$ are the dimensionalities of the corresponding spaces. If the dimensionality of the map is low enough $(s \leq 3)$, the so-called ``map points'' can be visualised in a scatter plot. The aim is to preserve latent features of the source data in the reduced representation.

Manifold learning is a nonlinear approach to dimensionality reduction which assumes that the source data lies on, or close to, some low-dimensional manifold embedded within the original high-dimensional space \cite{lee2007nonlinear,lin2008riemannian}. The idea is to produce a low-dimensional representation of the source data by deducing its projected layout across the manifold - this principle is illustrated in Fig.~\ref{fig:dim_red}. This has advantages over linear methods such as principal component analysis (PCA) \cite{hotelling1933analysis} in that it allows the manifold onto which the data is projected to adopt a folded geometry within the original space, making no assumptions about this geometry. The faithfulness of the projection depends on how close the source data is to actually lying on some embedded manifold of the desired dimension, how good the cover of the source data across this manifold is, and how well the given algorithm is able to capture the key features of the manifold from the spread of the source data.

There are many manifold learning algorithms, generally differing in the geometrical properties of the underlying manifold that they seek to preserve. Some aim to maintain distances or similarities across the entire data-set, e.\,g.\ Isomap \cite{tenenbaum2000global} or kernel PCA \cite{scholkopf1998nonlinear}. Others prioritise the preservation of local proximities, e.\,g.\ Sammon mapping \cite{sammon1969nonlinear}, Locally Linear Embedding (LLE) \cite{roweis2000nonlinear}, or Laplacian Eigenmaps \cite{belkin2002laplacian}. For source data that is nonlinear, it is usually more important to keep the low-dimensional representations of similar points close together, rather than trying to accurately reproduce a given metric at all length-scales \cite{maaten2008visualizing}. The t-SNE \cite{maaten2008visualizing}, LargeVis \cite{tang2016visualizing}, and UMAP \cite{mcinnes2018umap1} algorithms also follow this principle, but are able to patch together the local representations in such a way that global features are captured more faithfully than with other local methods \cite{maaten2008visualizing,mcinnes2018umap1}. These approaches proceed by constructing a weighted graph describing the structure of the source data, then projecting this into a space of fewer dimensions through the optimisation of a non-convex cost function. SHEAP is formulated within the same framework, the outline of which is provided next, followed by a discussion of the motivation for the development of SHEAP.

\subsection{Manifold learning through weighted graph matching}\label{framework}

A weighted graph is constructed, consisting of similarities $p_{ij}$ between each pair of data-points $i$ and $j$, corresponding to a measure of how close they are in the original space $\mathbb{R}^D$ according to some distance metric $d : X \times X \rightarrow \mathbb{R}$. It is usually beneficial to the speed and scalability of a given algorithm to only compute weights to the points in the region local to each data-point (e.\,g.\ the $k$ nearest neighbours), with all other weights being set to zero - this is exploited in LargeVis \cite{tang2016visualizing}, UMAP \cite{mcinnes2018umap1}, and accelerated (approximate) implementations of t-SNE \cite{van2013barnes,linderman2017efficient}. An initial distribution of $N$ data-points is then produced in a space $\mathbb{R}^s$ of many fewer dimensions (typically 2 or 3). This may be generated at random, possibly according to some bias(es) and/or constraint(s), or through the application of some more primitive algorithm for dimensionality reduction, such as PCA or random projection (see Appendix~\ref{random_projection}). Another graph of weights $q_{ij}$ is then constructed for the nodes defined by these map points, $Y$, possibly taking a different functional form to those defined between the high-dimensional points. The low-dimensional representation of the data is then computed by adjusting the layout of the map points such that the faithfulness with which the weights $q_{ij}$ model each $p_{ij}$ is optimised. This is achieved by minimising some cost function $C(Y;X)$ designed to penalise any mismatch. For a non-convex cost function, local optimisation schemes are necessary. Since there is usually some element of randomness in the initialisation of the map points and/or the optimisation of the cost function, different local minima can be obtained. Hence, these methods are considered stochastic. Note that the resulting projection is not associated with any physically meaningful set of axes (unlike PCA, for example) - it is only the relative positioning and clustering of the mapped data that is significant.

Because there are numerous stages at which there is scope for different possible and justifiable algorithmic choices, many algorithms are possible. These choices include: how the weighted graphs describing the connectivity in the high and low dimensions are constructed, what cost function is optimised, how the map points are initialised, and what optimisation scheme is used.

\subsection{Motivation for SHEAP}

t-SNE has become perhaps the most widely used manifold learning algorithm for visualisation. However, in many test cases the more recent methods LargeVis (2016) and UMAP (2018) appear to offer improvements in terms of the quality of the visualisation, run-time speed, and scalability \cite{tang2016visualizing,mcinnes2018umap1}. Approximate implementations, such as Barnes-Hut-SNE \cite{van2013barnes} and Fast Fourier Transform accelerated Interpolation-based t-SNE (FIt-SNE) \cite{linderman2017efficient} dramatically accelerate the computation of t-SNE, though these too appear inferior to UMAP for many test data sets \cite{mcinnes2018umap1}.

Here we focus on the relative visualisation qualities of these algorithms, leaving speed and scaling to future work. As demonstrated in Ref. \cite{mcinnes2018umap1}, UMAP produces a local structure that is comparable to the projections of t-SNE (and LargeVis), but appears able to provide a more faithful global representation. This is likely a result of a better cost function - UMAP's choice provides a more sophisticated treatment of the repulsive contribution to the interaction between map points during optimisation by introducing greater dependency on the layout of the source data. This is addressed further in Section \ref{SHEAP}, and discussed in detail in Appendix~\ref{UMAP_tsne}.

However, UMAP does not truly optimise the cost function it defines. Rather, the algorithm relies on its specific choice for the optimisation procedure, an approximate implementation of stochastic gradient descent (SGD) which uses probabilistic edge sampling and negative sampling \cite{mcinnes2018umap1,tang2016visualizing}, to de-exaggerate the repulsive part of the cost function gradient. This is somewhat unsatisfactory, and limits the flexibility of the algorithm. For example, UMAP's reliance on SGD makes it incompatible with our approach to incorporating basin volumes into the maps - this involves representing the structures in the map as hard spheres/circles, rather than single points (see Section \ref{hard_spheres}), and is straightforward to incorporate within t-SNE. Thus, we develop SHEAP, a hybrid of UMAP and t-SNE which captures some of the advancements made by UMAP, whilst still producing an algorithm that: (1) truly optimises a well defined and theoretically justified cost function, and (2) is compatible with the hard sphere scheme we introduce in Section \ref{hard_spheres}.

\section{Stochastic Hyperspace Embedding And Projection (SHEAP)}\label{SHEAP}

In SHEAP, the weighted graph in the high-dimensional space is constructed according to the same scheme as in t-SNE. Following Ref.~\cite{maaten2008visualizing}, the weight $p_{ij}$ between source data-points $i$ and $j$ is defined as
\begin{equation}
    p_{ij} = \frac{p_{j|i} + p_{i|j}}{2 N} \; ,
\label{pij_symmetrise}
\end{equation}
where
\begin{equation}
    p_{j|i} = \frac{\exp \left( - d(\mathbf{x}_i,\mathbf{x}_j)^2 / 2 \sigma_i^2 \right)}{\sum_{k \neq i} \exp \left( - d(\mathbf{x}_i,\mathbf{x}_k)^2 / 2 \sigma_i^2 \right)} \; .
\label{pij_gaussian}
\end{equation}
Here, $d(\mathbf{x}_i,\mathbf{x}_j)$ is some metric for computing the distance between $\mathbf{x}_i$ and $\mathbf{x}_j$ in the high-dimensional space. t-SNE uses the $\ell^2$-norm: $d(\mathbf{x}_i,\mathbf{x}_j) = || \mathbf{x}_i - \mathbf{x}_j ||_2$. The value of $\sigma_i$ is chosen such that the so called perplexity of $P_i$ takes some desired value, where $P_i$ is the conditional probability of all other data-points given $\mathbf{x}_i$. The perplexity of $P_i$, $\text{Perp}(P_i)$, is computed as
\begin{equation}
    \text{Perp}(P_i) = 2^{H(P_i)} \; ,
\end{equation}
where
\begin{equation}
    H(P_i) = - \sum_j p_{j|i} \log_2 p_{{j|i}} \; .
\end{equation}

The perplexity $\text{Perp}(P_i)$ can be interpreted as a smooth measure of the effective number of neighbours of $i$, and is set to the same value for all data-points. It can be thought of as defining how far each point ``looks out'' to describe its local environment. A smaller perplexity favours a faithful representation of the local connectivity of the data, whilst a larger value favours the global distribution. A ``good'' choice for this parameter results from a compromise between these two competing goals. A potential drawback arises in that the perplexity is a global parameter; there may be no single value that is suitable for all source data points and their local regions. It may be the case that a range of perplexities must be explored in order to gain the greatest insight from these projections. Which single perplexity value leads to the the most complete picture can then be assessed.

Given a user input perplexity, each $\sigma_i$ is computed using a straightforward binary search. Because the distributions of any two data-points $i$ and $j$ will likely require different $\sigma$'s in order to correspond to the same perplexity, the conditional probabilities $p_{j|i}$ and $p_{i|j}$ will in general be different. To obtain a symmetric measure for the weight $p_{ij}$ between the pair of data-points, the joint probability of $p_{j|i}$ and $p_{i|j}$ is computed, as in Eqn.~(\ref{pij_symmetrise}).

In the low-dimensional space, the weight between $i$ and $j$ is computed according to a Student t-distribution, similarly to in t-SNE, but using a different normalisation:
\begin{equation}
    q_{ij} = \frac{\left( 1 + || \mathbf{y}_i - \mathbf{y}_j ||_2^2 \right)^{-1}}{N(N-1)} \; .
\end{equation}
As outlined in Ref.~\cite{maaten2008visualizing}, the use of a heavier-tailed distribution for the low-dimensional weights allows moderate distances in the high-dimensional space to be faithfully represented by much larger distances in the projection, encouraging (larger) gaps to form in the low-dimensional map between the natural clusters present in the source data, alleviating the so called crowding problem.

The key difference of the SHEAP algorithm from t-SNE is the choice of the cost function:
\begin{equation}
    C = \sum_{i,j} p_{ij} \log \left( \frac{p_{ij}}{q_{ij}} \right) + (1-p_{ij}) \log \left( \frac{1-p_{ij}}{1-q_{ij}} \right) \; .
    \label{ce_UMAP}
\end{equation}
This is the fuzzy set cross-entropy between the two sets of weights, as in UMAP, rather than KL divergence (Eqn.~\ref{KL_tsne}), used in t-SNE. Recall that we attributed the improved global structure in UMAP's projections over t-SNE's to this choice of the cost function, due to a better treatment of the repulsive contribution to the interaction between the map points (see Appendix~\ref{UMAP_tsne}). In t-SNE, this repulsion arises due to the constraint that all of the weights $q_{ij}$ sum to 1, providing a uniform repulsion between the map points. In UMAP, the repulsive contribution arises directly from the second, additional term in the cost function, and is much more dependent on the structure of the source data. This also allows us to discard the computationally costly normalisation appearing in the weights $q_{ij}$ in t-SNE, which must be recomputed at every step during optimisation. The constant normalisation factor that replaces this here is chosen such that weights $q_{ij}$ are consistent with the weights $p_{ij}$ in limit that all data-points lie on top of one another in both the source data and the projection, giving $p_{ij}=q_{ij}=\frac{1}{N(N-1)}$ for all $i$ and $j$.

Given the above choices, the cost function gradient in SHEAP is given by:
\begin{equation}
    \frac{\partial C}{\partial \mathbf{y}_i} = 4 N(N-1) \sum_j \left( p_{ij} - q_{ij} \frac{(1-p_{ij})}{(1-q_{ij})} \right) q_{ij} (\mathbf{y}_i - \mathbf{y}_j) \; .
\end{equation}

Optimisation of the low-dimensional layout is achieved using the two-point steepest descent (TPSD) scheme of Barzilai and Borwein \cite{barzilai1988two}, and the option of early exaggeration is also implemented, as in t-SNE \cite{maaten2008visualizing,linderman2017clustering}. We can replace the use of early exaggeration with an initial use of SGD, though this does not appear to improve performance.

The layout of map points is initialised with an algorithm for random projection - see Appendix~\ref{random_projection}. This tremendously accelerates the cost function optimisation, relative to initialising with a random (confined) distribution of points.

Here we produce low-dimensional depictions of energy landscapes by applying this algorithm to sets of structures in the relevant configuration space. Primarily, we consider data-sets containing only local minima, obtained through random structure searching (RSS) \cite{pickard2007structure,pickard2011ab} to promote broad sampling of the space. In Section \ref{results} we demonstrate that this is sufficient for capturing the PES features of interest in the context of structure prediction. In Section \ref{LJ38} we also explore the projection of unrelaxed structures obtained through random sampling.

\subsection{Representing structural data}

The configuration of a system of atoms can be represented in terms of a structure descriptor - a vector of real valued functions of the atomic coordinates which describes their relative arrangement \cite{bartok2013representing}. To allow comparisons between structures, a good descriptor is invariant with respect to permutations of equivalent atoms, as well as global translations, rotations and reflections, and uniquely determines the structure up to these symmetries.

Two descriptors are considered here. The first is a sorted list of the pairwise distances between atoms. It is known that this is not a complete descriptor \cite{bartok2013representing}. However, for the simple systems of LJ particles considered (for which the energy depends only on pairwise distances), we find this to capture a sufficient proportion of the structural information/variation.

The second descriptor considered is the smooth overlap of atomic positions (SOAP) \cite{bartok2013representing}. SOAP encodes relative nuclear geometry in the neighbourhood of a given atom using an atomic density field, represented by 3D atom-centred Gaussian functions expanded in a basis of spherical harmonics and radial basis functions. A global SOAP descriptor for a given structure can be computed by taking the average SOAP fingerprint between all atomic environments \cite{de2016comparing}. When multiple species are present, one can either compute a single average incorporating all species, or average separately for each species and then concatenate the species-specific averaged vectors \cite{cheng2020mapping} - here we adopt the former approach. Recent work by Cheng et al.~\cite{cheng2020mapping} has found success in pairing SOAP descriptors with kernel PCA to produce low-dimensional maps for a wide variety of structural data-sets.

The distance metric $d$ used here is the $\ell^2$-norm, as in t-SNE \cite{maaten2008visualizing}. This implicitly assumes that the source data is locally linear on the underlying manifold \cite{maaten2008visualizing}. The SHEAP map produced from a given set of structures is expected to depend on both the chosen descriptor and distance metric. Here we explore the the former through a comparison of SHEAP maps produced from sorted lists of distances and from SOAP, for a single LJ system. An investigation into other distance metrics will be the subject of future work.

\subsection{Basin volumes conserved using hard spheres}\label{hard_spheres}

A shortcoming of using only the spread of local minima to depict a PES is that this gives no sense of how large each of the basins of attraction are. To address this, we extend the methodology outlined in Sections \ref{framework} and \ref{SHEAP} by representing each local minimum in the map as a hard $n$-sphere $(n = s-1)$, rather than a point, with volume (area in 2D) proportional to the (approximate) volume of its basin.

The number of times that structure $i$ is located in a search is labelled $c_i$. We consider structures to be equivalent if the distance between their descriptor vectors falls below an appropriate similarity threshold. For a suitably large search, these counts should give representative estimates for the relative volumes of the corresponding basins. We represent each structure in the map by a sphere of radius
\begin{equation}
    R_i = (c_i)^{\frac{1}{s}} R_0 \; ,
    \label{sphere_radius}
\end{equation}
such that the volume $V_i \propto R_i^s$ is proportional to the count $c_i$. The quantity $R_0$ is the minimum sphere radius appearing in the map, taken by any structure that has been found just once - the choice of this parameter is important, and is discussed in detail in Appendix~\ref{R_0}.

The hard spheres are incorporated into the framework outlined in Section \ref{framework} through the addition of an extra term to the cost function. This hard sphere potential is defined as:
\begin{equation}
    U_{\text{HS}} = \frac{1}{2} \gamma_{\text{HS}} \sum_i \sum_{j \neq i} \alpha_{ij}^2 \; ,
\end{equation}
where $\gamma_{\text{HS}}$ is the hard core strength, and
\begin{equation}
    \alpha_{ij} = \begin{cases}
         \frac{R_i + R_j - || \mathbf{y}_i - \mathbf{y}_j ||}{2} &\text{if $|| \mathbf{y}_i - \mathbf{y}_j || < R_i + R_j$}\\
         0 &\text{otherwise.}
    \end{cases}
\end{equation}
For a sufficiently large value of $\gamma_{\text{HS}}$, the repulsion between any overlapping spheres (when present) will be the dominant contribution to the gradient, meaning the layout will be forced to contain no overlaps.

The inclusion of the hard sphere potential from the start of the optimisation is problematic. The spheres become jammed as they try to move across one-another, particularly for densely packed spheres, resulting in convergence being drastically slow or not attainable at all. This can be avoided by allowing the spheres to access additional dimensions during the optimisation, using the geometry optimisation of structures from hyperspace (GOSH) method \cite{pickard2019hyperspatial}. Alternatively, an efficient solution is to initially run the optimisation without the hard sphere contribution, switching it on only when the algorithm is close to convergence. With this, convergence is more easily attained if the hard spheres are introduced at a fraction of their final value, and expanded over the course of a few hundred optimisation steps. In this scheme, the hard spheres have minimal impact on the optimisation, essentially just distorting the optimal layout once it has already been reached.

Again we point out that this approach is incompatible with the SGD optimisation scheme relied on by the UMAP algorithm - large overlaps between spheres are produced due to the omission of certain contributions to the gradient, resulting in instability in the optimisation. A similar problem is encountered with the use of hard spheres at the same time as early exaggeration, though this issue can straightforwardly be circumvented by ensuring that the hard sphere interaction is not switched on until after the exaggeration period has ended.

\subsection{Energy represented with colour}

Energy is depicted across our plots through the colouration of the circles/spheres representing each structure. In Sections \ref{results} we also demonstrate that insight can be gained by illustrating the variation of other quantities across the projected configuration space.

\section{Results}\label{results}

The results of applying SHEAP to various systems are now presented. First, we consider the PESs defined by LJ clusters, with pairwise interactions governed by:
\begin{equation}
v_{\text{LJ}}(r) = 4 \epsilon_{\text{LJ}} \left( \frac{\sigma^{12}_{\text{LJ}}}{r^{12}} - \frac{\sigma^6_{\text{LJ}}}{r^6} \right) \; ,
\end{equation}
where $\epsilon_{\text{LJ}}$ is the depth of the potential energy well for a given particle pair, and $\sigma_{\text{LJ}}$ is the finite distance at which the inter-particle potential is equal to zero. We set $\epsilon_{\text{LJ}} = 1$ and $\sigma_{\text{LJ}} = 2$ throughout. Clusters of $13$, $38$ and $55$ atoms are considered.

For each data-set of LJ clusters, structures are represented as a sorted list of all pairwise distances between atoms, unless explicitly stated otherwise. In Appendix~\ref{LJ13_inv} we also explore the use of a sorted list of inverse pairwise distances for $\text{LJ}_{13}$.

Following the study of these model systems, data-sets of realistic solid state systems, modelled by first-principles density-functional theory (DFT) calculations, are addressed.

The examples considered illustrate the applicability of dimensionality reduction by manifold learning to minima data obtained from RSS, and demonstrate the insight that can be gained from the resulting maps. A key finding is the intrinsic low-dimensionality in the distributions of minima across the corresponding PESs.

\subsubsection*{Computational details}

SHEAP is implemented in an OpenMP parallelised Fortran package, available under the GPL2 licence at \url{https://bitbucket.org/bshires/sheap/src/master/}. Also implemented in this package are the t-SNE algorithm, as well as a simplified version of UMAP - the qualitative performance of SHEAP on standard data-sets is compared against these in Appendix~\ref{standard_datasets}. The maps presented in this work are plotted with Ovito, the Open Visualization Tool \cite{stukowski2009visualization}. The structural data presented in each map is generated using the AIRSS structure prediction package \cite{pickard2006high,pickard2011ab}. SOAP vectors are computed using the implementation in ASAP \cite{cheng2020mapping}.

SHEAP is stochastic, so the final map differs for independent projections of the same source data-set, even for identical parameters. However, we find that for an appropriately chosen parameter set, distinct runs consistently provide qualitatively equivalent results in all cases - for each data-set we present the map for a single random seed.

\subsection{$\text{LJ}_{13}$}\label{LJ13}

\begin{figure}[t]
    \centering
    \includegraphics[width=\linewidth]{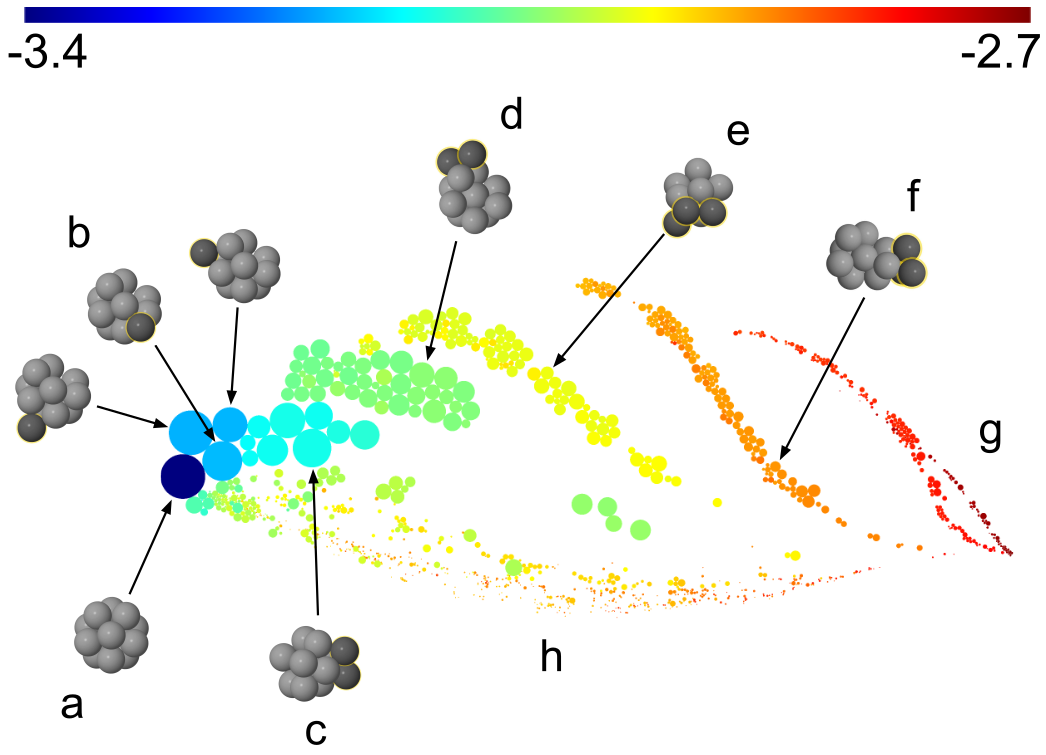}
    \caption{Labelled 2D SHEAP map produced from a set of $1{,}108$ distinct minima on the $\text{LJ}_{13}$ PES, located with RSS, described by sorted lists of all pairwise distances. The projection was made using a perplexity of 30, with a minimum sphere radius $R_0 = 0.005$. The circles representing each structure are coloured according to energy per atom in Lennard-Jones units; see colour bar at top of figure. Basin volumes are represented by the area of each circle. Labels (a)-(g) denote clusters of structures of similar energy and configuration; representative structures from each are shown. Structures in (h) are less obviously clustered, and tend to be comprised of less ordered packings of atoms.}
    \label{fig:LJ13_labelled_surface}
\end{figure}

\begin{figure}[t]
    \centering
    \includegraphics[width=\linewidth]{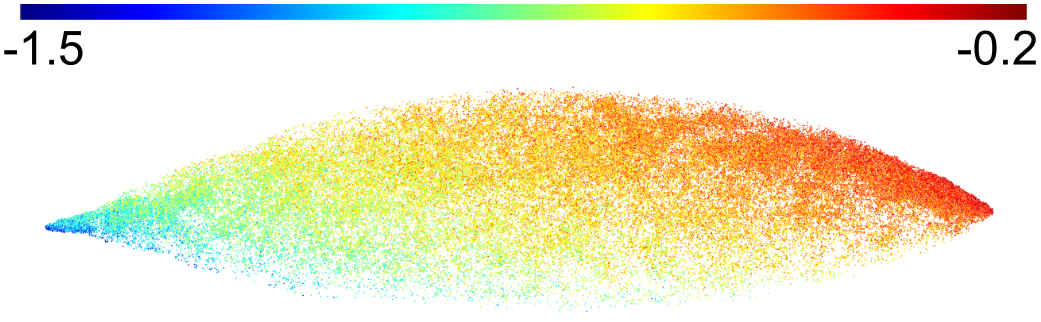}
    \caption{2D SHEAP map produced from $50{,}000$ unrelaxed, randomly sampled structures on the $\text{LJ}_{13}$ PES, described by sorted lists of all pairwise distances. The projection was made using a perplexity of 30, with a minimum sphere radius $R_0 = 0.01$. The circles representing each structure are coloured according to energy per atom in Lennard-Jones units; see colour bar at top of figure.}
    \label{fig:LJ13_unrelaxed}
\end{figure}

\begin{figure*}[t]
    \centering
    \includegraphics[width=0.9\linewidth]{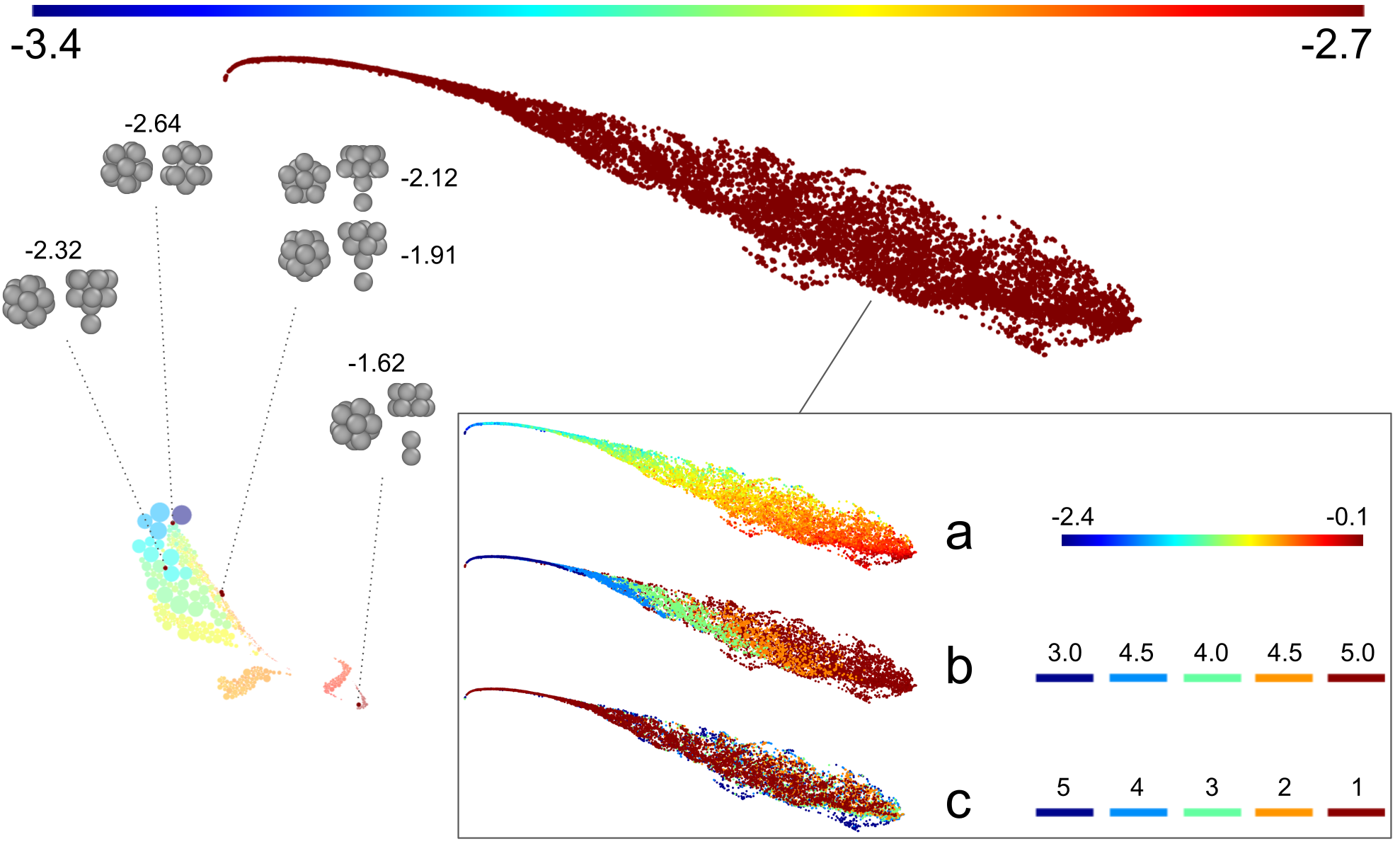}
    \caption{2D SHEAP map constructed from a data-set of minima and unrelaxed structures on the $\text{LJ}_{13}$ PES, described by sorted lists of all pairwise distances. The projection was made using a perplexity of 8. Circles in the main figure are coloured according to energy per atom in Lennard-Jones units, given by the colour bar at top of figure. Minima (all in the bottom-left cluster) are faded, with basin volumes represented by the area of each circle, using a minimum sphere radius $R_0 = 0.005$. Circles representing unrelaxed structures have been enlarged to highlight them. The figure inset provides different colourations of the top cluster, which contains only unrelaxed structues. Colouring is according to (a) energy per atom in Lennard-Jones units, (b) the confinement radius in Lennard-Jones units, and (c) the number of imposed symmetry operations - see corresponding colour bars to right of inset. The few unrelaxed structures that cluster with the minima (all possessing 5 symmetry operations) have been labelled with their energy per atom in Lennard-Jones units, alongside 2 perpendicular perspectives of their configuration.}
    \label{fig:LJ13_merge}
\end{figure*}

\begin{figure}[t]
    \centering
    \includegraphics[width=\linewidth]{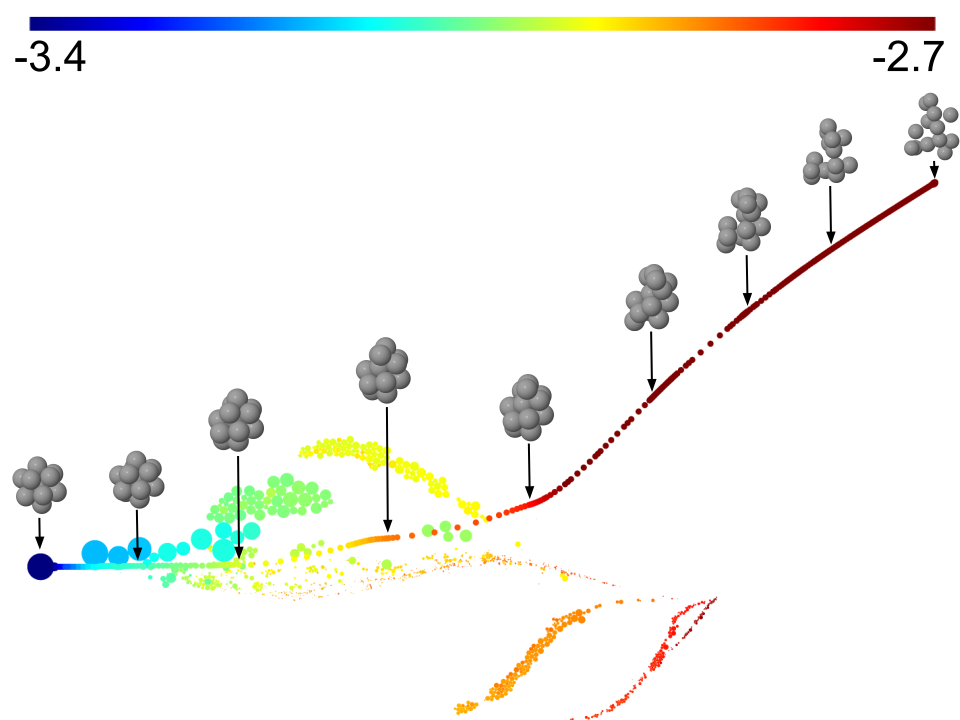}
    \caption{2D SHEAP map produced from a data-set consisting of local minima on the $\text{LJ}_{13}$ PES, plus selected structures visited along the pathway of steepest descent from an unrelaxed structure generated without symmetry constraint, down to the global minimum. Structures are described by sorted lists of all pairwise distances. The projection used a perplexity of 30. Circles are coloured according to energy per atom in Lennard-Jones units, given by the colour bar at top of figure. For the minima, basin volumes are represented by the area of each circle, with a minimum hard sphere radius $R_0 = 0.005$. Circles representing structures visited during the relaxation have been enlarged to highlight them.}
    \label{fig:LJ13_descent}
\end{figure}

$\text{LJ}_{13}$ has a relatively simple energy landscape. It is described as having a ``single funnel'' topology, meaning there is a general decrease in the energy of local minima with increasing structural similarity to the global minimum. This ground state structure is a filled 12-vertex icosahedron, and all local minima are within three rearrangements of it \cite{doye1999evolution}.

\subsubsection{Relaxed vs.~unrelaxed structures}

Fig.~\ref{fig:LJ13_labelled_surface} shows a 2D SHEAP map produced from local minima on the $\text{LJ}_{13}$ PES. The search, which comprised of $50{,}000$ samples, used a confinement sphere of radius $5.0~\sigma_\text{LJ}$ for the initial random structures, with an imposed minimum separation of $2.0~\sigma_\text{LJ}$. Structures were relaxed to nearby local minima using TPSD \cite{barzilai1988two}, resulting in a data-set of $1{,}108$ distinct minima.

The funnelled structure of the landscape is apparent, with minima energies correlating well with their distance from the global minimum (a) in the map. In addition, SHEAP has clustered together structures of similar energies and arrangements. For example, the structures labelled (b) all result from one external atom of the icosahedron being moved from its ground state position to a different surface site. Structures in the cluster labelled (c) result from two external atoms from the icosahedron being moved to different surface sites.

The success observed in the application of SHEAP to the data-set of minima is in stark contrast to the map presented in Fig.~\ref{fig:LJ13_unrelaxed}, which displays the 2D SHEAP projection of the $50{,}000$ unrelaxed structures. Fig.~\ref{fig:LJ13_unrelaxed} contains no obvious clustering. The structures are arranged in a featureless, elongated ``blob'', roughly ordered according to energy. Similar results are observed when projecting into 3D.

The disparity between Figs.~\ref{fig:LJ13_labelled_surface} and \ref{fig:LJ13_unrelaxed} is striking. It suggests that the PES itself is truly high dimensional, and one cannot produce a meaningful representation of the landscape from a projection of points drawn randomly from it. In the language of Sec.~\ref{dim_red}, the randomly sampled points do not appear to lie on or close to any low-dimensional manifold. The distribution of minima on this surface, on the other hand, does appear to have some inherent low dimensionality, according to the choice of structure descriptor and distance metric.

\subsubsection{Projecting relaxed and unrelaxed structures together}

The question then is, where do these unrelaxed structures lie relative to the minima? One might intuitively predict that they should fall into the gaps between the minima, but this is not what is observed using SHEAP.

Fig.~\ref{fig:LJ13_merge} shows a 2D SHEAP map constructed from a data-set comprised of local minima and unrelaxed structures on the $\text{LJ}_{13}$ PES. The data-set contains the same relaxed structures that are presented in Fig.~\ref{fig:LJ13_labelled_surface}, as well as a series of unrelaxed random structures generated according to different constraints. These are: (1) a specified number of symmetry operations, from 2 up to 5, with a fixed confinement sphere radius of $5.0~\sigma_{\text{LJ}}$, and (2) no symmetry constraints, with varying confinement sphere radius, from $3.0~\sigma_{\text{LJ}}$ up to $5.0~\sigma_{\text{LJ}}$ in increments of $0.5~\sigma_{\text{LJ}}$. There are $1{,}000$ structures included for each set of constraints. Note that the distribution of minima is distorted by the addition of the unrelaxed structures. We attribute this to a reduced sensitivity in the relations between the minima, due to having to represent similarities to structures not on the manifold occupied by the minima. A lower perplexity was found to alleviate this somewhat.

The majority of unrelaxed structures lie well away from the minima in a single, elongated cluster. As with Fig.~\ref{fig:LJ13_unrelaxed}, these unrelaxed structures are arranged roughly according to energy. Here, we also see that denser packings (smaller confinement sphere radius) tend to lie towards the low-energy end of the cluster, closest to the minima, whilst the higher-symmetry packings tend to lie towards the edges. Interestingly, a few of the unrelaxed structures actually cluster with the minima, despite being of significantly higher energies. Each of these structures is found to possess 5 symmetry operations, and all contain at least 11 atoms that are close to the ideal icosahedral close-packing. This result demonstrates one aspect of the AIRSS philosophy - that sensible random structure generation does much of the work in efficiently locating local minima.

\subsubsection{Projecting a steepest descent pathway}

In order to reveal the connection between the pre-relaxed structures and the central layout of local minima in Fig.~\ref{fig:LJ13_merge}, a SHEAP map was produced for a source data-set consisting of the same set of minima, plus selected structures visited along the pathway of steepest descent from an unrelaxed structure generated without symmetry constraint, down to the global minimum. The resulting projection is provided in Fig~\ref{fig:LJ13_descent}.

We observe that this pathway traverses the space in the map between the unrelaxed structures and the minima, and passes by many local minima on its way to the icosahedral ground-state. The presence of the descent pathway leads to slight distortions in the distribution of minima in the map, relative to Fig.~\ref{fig:LJ13_labelled_surface}, likely due to over-sampling of the regions on the PES visited during the relaxation.

These SHEAP maps for $\text{LJ}_{13}$ provide an interesting perspective on the structure of its energy landscape. They suggest that the surface consists of clusters of minima which all sit very close to one another, with the unrelaxed structures predominantly residing outside the region occupied by these minima. Furthermore, they imply the existence some inherent low dimensionality in the distribution of these minima across the surface.

\subsection{$\text{LJ}_{38}$}\label{LJ38}

\begin{figure*}[hbtp]
    \centering
    \includegraphics[width=0.85\linewidth]{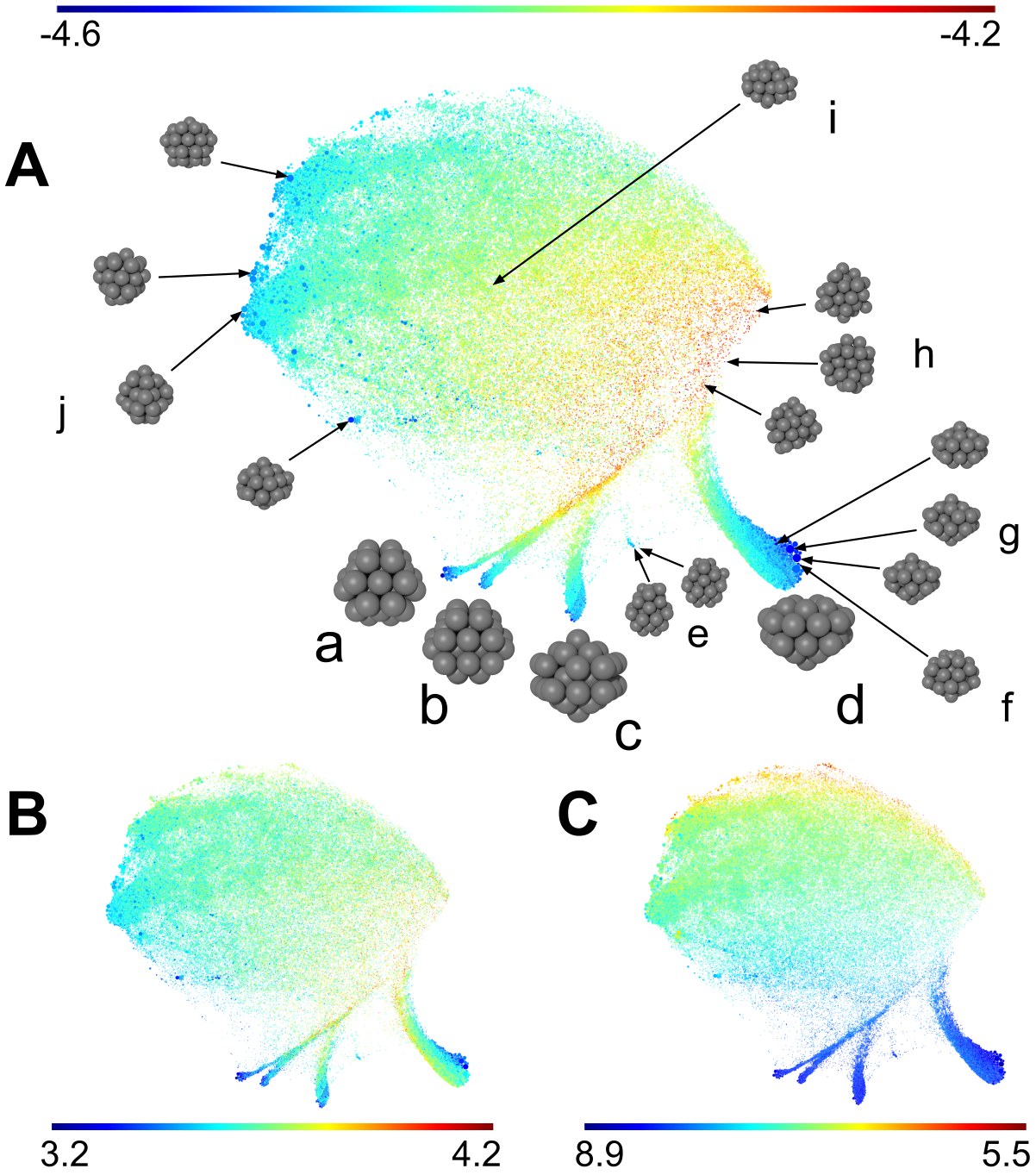}
    \caption{(A) shows a 2D SHEAP map for $89{,}464$ minima on the $\text{LJ}_{38}$ PES, generated with RSS, described by sorted lists of all pairwise distances. The projection was made using a perplexity of 30, with a minimum sphere radius $R_0 = 0.008$. The circles representing each structure are coloured according to energy per atom in Lennard-Jones units; see colour bar at top of figure. Basin volumes are represented by the area of each circle. (a)-(d) label proposed funnels in the $\text{LJ}_{38}$ PES; structures shown alongside correspond to the lowest energy minima residing in each. (a) contains the octahedral global minimum; (b) and (c) could be considered sub-funnels of this. (d) contains the icosahedral lowest meta-stable structure. (e) labels a small island of structures occupying the space between (a)-(c) and (d). (f) and (g) label low-energy minima residing in funnel (d), and are closely related to icosahedral structure. (h), (i) and (j) label representative structures from the large, ``brain-like'' cluster of minima which contains the majority of structures - these are predominantly of low-symmetry. (B) and (C) depict the same SHEAP map as in (A), but coloured according to different physical parameters. Circles in (B) are coloured according the convex hull volume of each cluster. Circles in (C) are coloured according to number of contacts per atom in each cluster (see main text for details). Corresponding colour bars are provided below these sub-figures.}
    \label{fig:LJ38_labelled_surface}
\end{figure*}

\begin{figure}[t]
    \centering
    \includegraphics[width=0.9\linewidth]{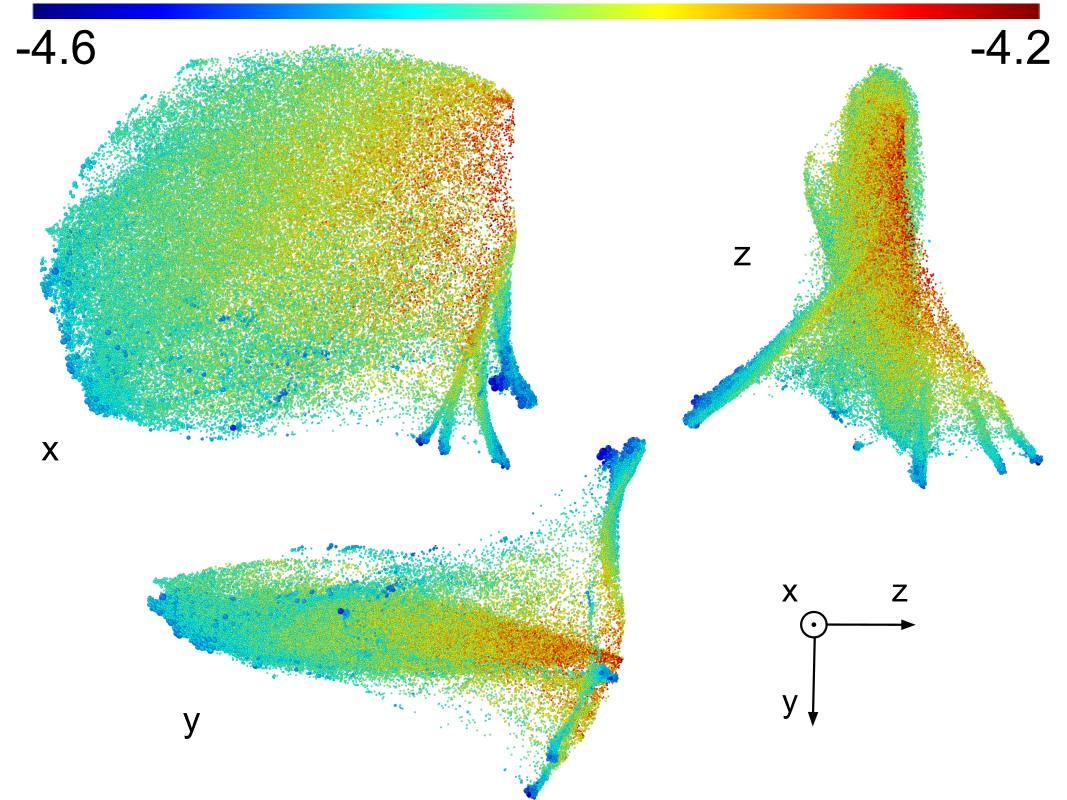}
    \caption{Three perpendicular perspectives of a 3D SHEAP map produced from the same set of $\text{LJ}_{38}$ minima that are presented in Fig.~\ref{fig:LJ38_labelled_surface}. The projection was made using a perplexity of 30, with a minimum sphere radius $R_0 = 0.008$. The spheres representing each structure are coloured according to energy per atom in Lennard-Jones units; see colour bar at top of figure. Basin volumes are represented by the volume of each sphere. Each viewpoint looks down one of the axes displayed at the bottom right of the figure; the axes have no other physical meaning.}
    \label{fig:LJ38_3D_perspectives}
\end{figure}

\begin{figure*}[t]
    \centering
    \includegraphics[width=0.95\linewidth]{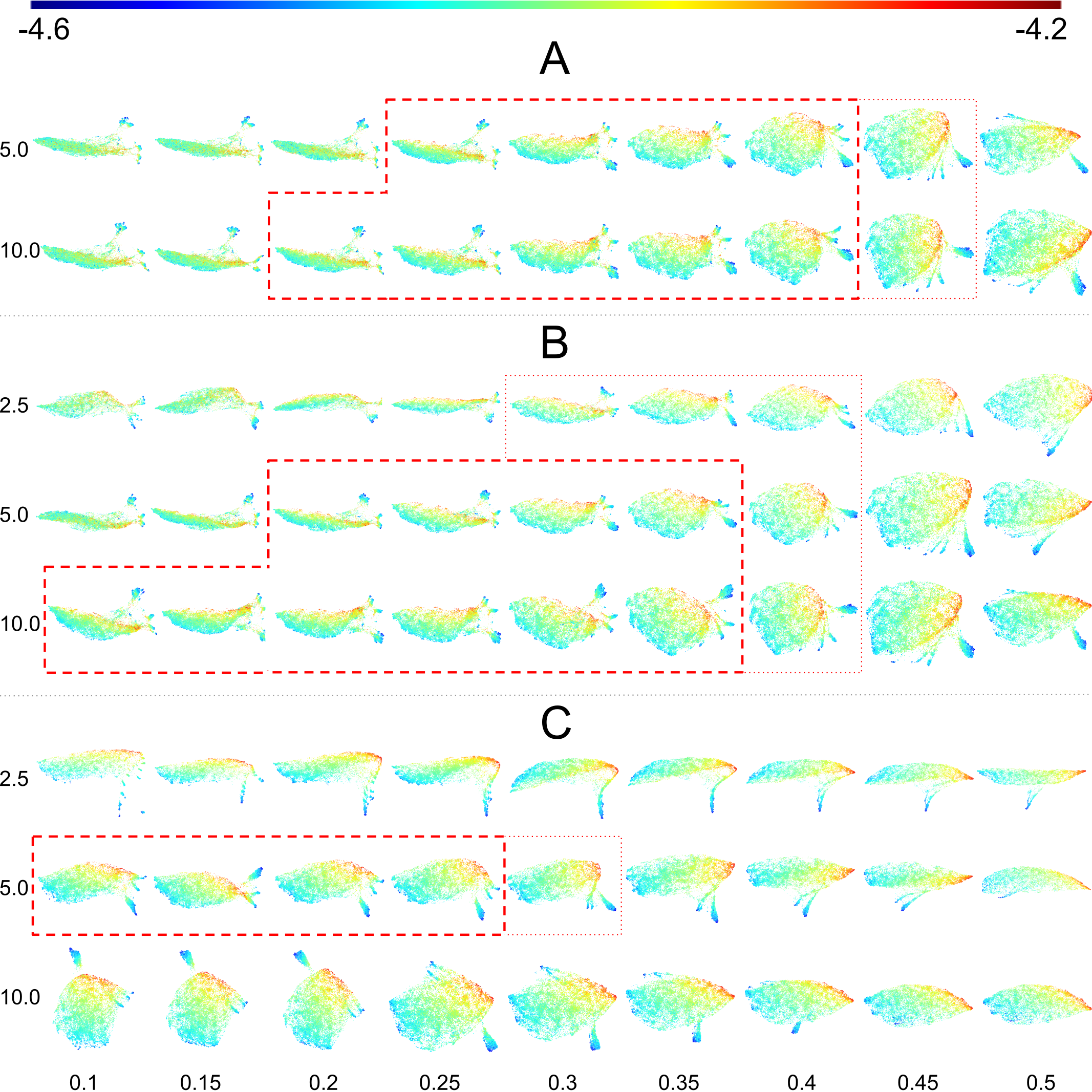}
    \caption{2D SHEAP maps for $10{,}000$ minima (prior to similarity check) on the $\text{LJ}_{38}$ PES, described using SOAP with various parameters. (A), (B), and (C) show results for $n_{\text{max}}, l_{\text{max}} = 15, 9$, $12, 6$, and $8, 4$, respectively. $r_{\text{cut}}$ is consistent across each row, values given on the far left. $\sigma$ is consistent across each column, values given at the bottom. Projections were made using a perplexity of 15, with $R_0$ computed by Eqn.~\ref{R_0}. The circles representing each structure are coloured according to energy per atom in Lennard-Jones units; see colour bar at top of figure. Basin volumes are represented by the area of each circle. Maps in the bold, dashed, red perimeters represent ``suitable'' SOAP parameter choices, producing the same key features as Fig.~\ref{fig:LJ38_labelled_surface}. Maps in the faded, dashed, red perimeters also show these features, but less clearly.}
    \label{fig:LJ38_soap_params}
\end{figure*}

\begin{figure*}[t]
    \centering
    \includegraphics[width=0.95\linewidth]{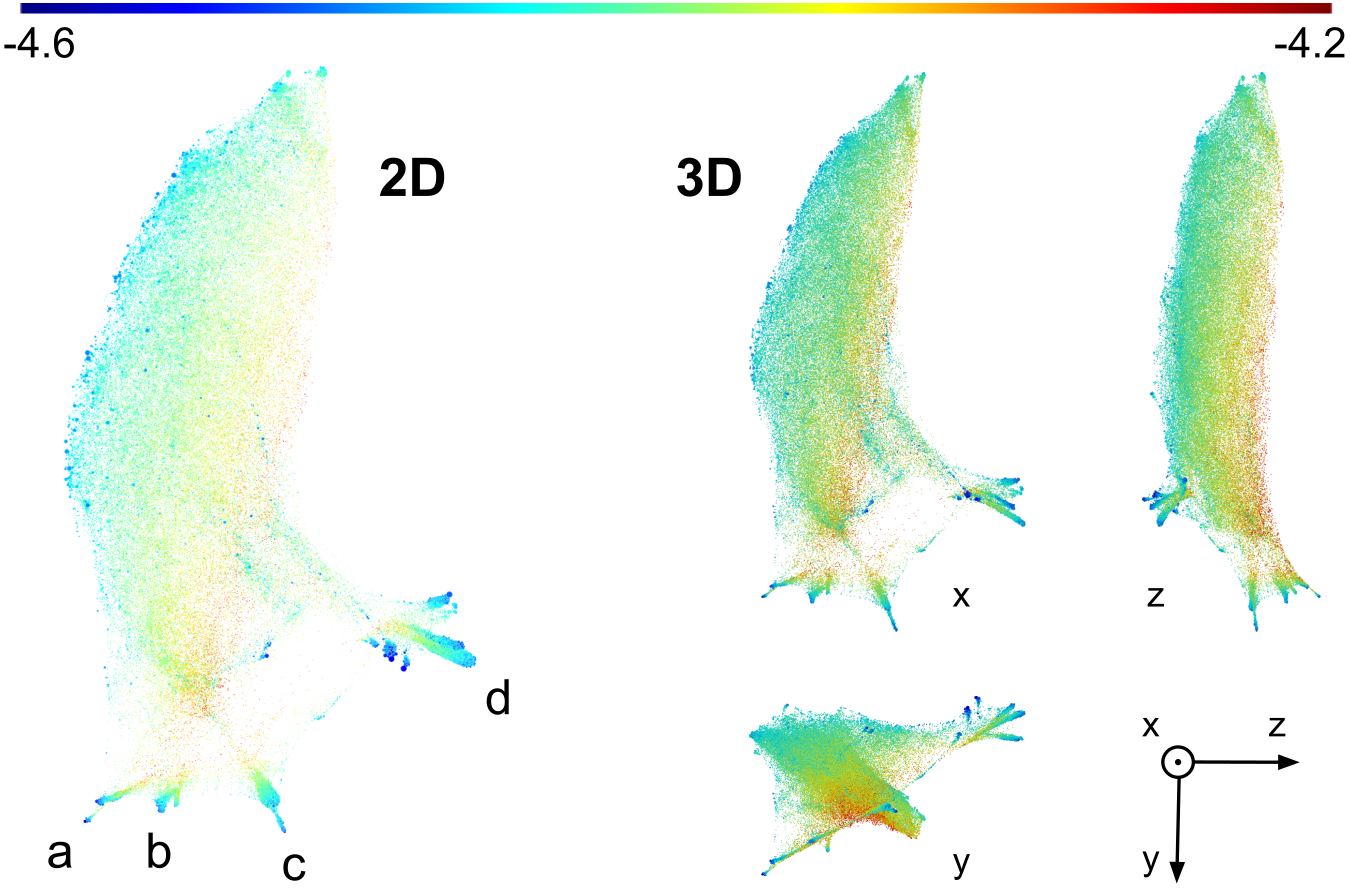}
    \caption{2D and 3D SHEAP maps for produced from the same set of $\text{LJ}_{38}$ minima that are presented in Fig.~\ref{fig:LJ38_labelled_surface}, described using SOAP with parameters $r_{\text{cut}} = 10.0$, $n_{\text{max}},l_{\text{max}} = 15, 9$, and $\sigma = 0.3$. The projections were made using a perplexity of 30, with a minimum sphere radius $R_0 = 0.006$ for the 2D map, and $R_0 = 0.01$ for 3D. The circles/spheres representing each structure are coloured according to energy per atom in Lennard-Jones units; see colour bar at top of figure. Basin volumes are represented by the area/volume of each sphere/circle. Each viewpoint of 3D map looks down one of the axes displayed, as labelled; the axes have no other physical meaning. (a)-(d) in the 2D map label proposed funnels in the $\text{LJ}_{38}$ PES, analogous to those labelled in Fig.~\ref{fig:LJ38_labelled_surface}, with the same low-energy minima residing at the ends.}
    \label{fig:LJ38_soap_big}
\end{figure*}

In previous work, $\text{LJ}_{38}$ has been described as having an archetypal double funnel energy landscape; this is depicted clearly in disconnectivity graphs, such as in Ref.~\cite{wales2015perspective}. The 38 atom system stands out in the context of LJ clusters due to its octahedral global minimum - most other cluster sizes exhibit ground-states based on icosahedra \cite{leary1997global}. Whilst the octahedron is the lowest energy arrangement for $\text{LJ}_{38}$, there is an icosahedral meta-stable configuration that is only very slightly higher in energy. These two minima reside at the bottom of the two separate ``funnels'' in the landscape, with the funnel corresponding to the icosahedral structure being much broader than that corresponding to the global minimum, i.\,e.\ containing many more minima \cite{neirotti2000phase}. For this reason, $\text{LJ}_{38}$ is well known to be a relatively difficult test for structure searching algorithms.

\subsubsection{Two-dimensional map of minima data}

Fig.~\ref{fig:LJ38_labelled_surface} (A) shows a 2D SHEAP map produced from minima on the $\text{LJ}_{38}$ PES. The search consisted of $100{,}800$ samples, and used the same parameters and relaxation scheme as for the 13 atom system, resulting in a data-set of $89{,}464$ distinct minima.

$\text{LJ}_{38}$ has many more degrees of freedom than $\text{LJ}_{13}$, leading to a far more complex landscape topology, with many more local minima. This is reflected in the searches, with almost 90 times more distinct minima being located for the larger cluster from little over twice as many samples, and a much lower disparity between the largest and smallest basin volumes. The map for $\text{LJ}_{38}$ exhibits greater connectivity than that for $\text{LJ}_{13}$, with the former showing much less clear-cut clustering of structures, and a more continuous variation in the energy. We attribute this to the vastly greater number of minima compared to $\text{LJ}_{13}$, with the variation between minima being much more incremental for the larger system. For example, starting with the global minimum and moving one atom from its ground state position to a surface site leads to a relatively much greater deviation from the ground state structure for $\text{LJ}_{13}$ than for $\text{LJ}_{38}$.

The standout features in Fig.~\ref{fig:LJ38_labelled_surface} (A) are the large ``brain-like'' cluster of minima (mostly low-symmetry, amorphous structures), and the strands of minima leading away from this with progressively decreasing energies. With regards to the latter, there is an obvious connection to be drawn to the supposed funnels. We find that the octahedral global minimum and the icosahedral lowest meta-stable minimum reside at the ends of two of these different strands, labelled (a) and (d), respectively. Preceding them are closely related structures, e.\,g.\ those labelled (g), which all result from a single external atom of the icosahedral structure being relocated to a different surface site. Satisfyingly, the existence of these ``funnels'' has arisen purely from an analysis of the minima, through the structural information captured in the descriptor vectors. In the construction of the corresponding disconnectivity graph, funnelling is inferred through an examination of transition state data, and an evaluation of minima connectivity in the context of minimum energy pathways via these transition states. Although, note that one can construct approximate disconnectivity graphs based solely on the structural relationships between minima (using fingerprint distances), circumventing the expensive transition state calculations by computing estimates for the energy barriers in the system \cite{schaefer2016computationally}.

The following question arises: what are the remaining branches of structures appearing in the SHEAP map? In Fig.~\ref{fig:LJ38_labelled_surface}, the most apparent of these additional strands of structures are labelled (b) and (c), alongside illustrations of the lowest energy structure residing in each. We propose these to be previously unreported funnels in the $\text{LJ}_{38}$ PES, both of which could potentially be considered sub-funnels of a super-funnel also containing (a) - they appear to be part of the same branch, eventually separating off to comprise independent funnels.

\subsubsection{Other colour schemes}

To better interpret the layout of structures across the map, two further schemes for colouring the circles are considered, in addition to the energy per atom. These are the convex hull volume of the structure per atom, and the number of contacts in each structure per atom, provided in Figs.~\ref{fig:LJ38_labelled_surface} (B) and (C), respectively.

The convex hull volume shows a very similar distribution to the energy, both in the main cluster of structures (albeit with slightly less variation), and in the funnels. We attribute this to the LJ energy being entirely dependent on the pairwise distances between atoms; there are no 3-body or higher terms. Hence, the energy is expected to be closely correlated to the packing density within the structures, which in turn should correlate well with the map position.

In order to produce Fig.~\ref{fig:LJ38_labelled_surface} (C), a contact is defined as any pairwise separation that is less than $2.3~\sigma_{\text{LJ}}$, $2.5\%$ above the minimum separation in the LJ pair potential. We observe that the funnels in the map correspond to regions of very high numbers of contacts per atom, and the number of contacts steadily reduces with increasing separation from these funnels, roughly from bottom to top of the main cluster.

\subsubsection{Projecting into the third dimension}

So far SHEAP maps have only been presented for two map dimensions. However, we find that often there is a non-negligible improvement in the quality of visualisation by accessing an extra dimension and projecting into 3D. To illustrate this, in Fig.~\ref{fig:LJ38_3D_perspectives} we present three perpendicular perspectives of a 3D SHEAP map produced for the same data-set of $\text{LJ}_{38}$ minima. Note that the axis system is arbitrary.

The 3D map is qualitatively very similar to that in 2D; the 2D map looks almost identical to a projection of the 3D distribution onto the x-y plane. However, one can see from the other two perspectives that this is not the full picture, and there are features protruding out of this plane that are not represented faithfully in 2D. Most notably, the separation between the icosahedral funnel (d) and the 3 other large (sub-) funnels (a)-(c) is more pronounced with the extra dimension. The classification of (a)-(c) as sub-funnels of a single, larger ``super-funnel'' is also more obvious in this projection.

\subsubsection{Using a different descriptor: SOAP}

The distance based descriptor employed so far is appropriate for LJ systems, since their energy is determined entirely by pairwise separations. However, the picture of an energy landscape obtained with SHEAP should be qualitatively robust to the choice of descriptor, as long as the relevant structural information is captured. To test this, we compare the above results for $\text{LJ}_{38}$, which use sorted lists of distances, with the use of SOAP vectors \cite{bartok2013representing}.

Unlike the sorted list of distances (for which the only parameter is the cut-off distance), the SOAP vector for a given structure is dependant on many parameter choices: $r_{\text{cut}}$ specifies the distance beyond which neighbour atoms are not taken into account, $n_{\text{max}}$ specifies the number of radial basis functions to be used for each local descriptor, $l_{\text{max}}$ specifies the maximum degree of the spherical harmonics, and $\sigma$ specifies the standard deviation of the Gaussians. This level of freedom can be beneficial, allowing the descriptor to be tailored to better suit a given system, but it demands more input on the part of the user. The choice of radial basis functions is also important, but is not addressed in detail here. By default, ASAP uses spherical Gaussian type orbitals \cite{jager2018machine}, rather than the original polynomial radial basis set \cite{bartok2013representing}, allowing much faster analytic computation \cite{himanen2020dscribe}.

Fig.~\ref{fig:LJ38_soap_params} shows 2D SHEAP maps for a subset of the $\text{LJ}_{38}$ data-set studied in Section \ref{LJ38}, described by SOAP vectors using a range of parameters. A reduced data-set ($10{,}000$ structures) is used to lessen the computational cost.

ASAP permits up to $15$ radial and $9$ spherical harmonic SOAP basis functions. Fig.~\ref{fig:LJ38_soap_params} (A) illustrates the effect of varying $\sigma$ and $r_{\text{cut}}$ for these values of $n_{\text{max}}$ and $l_{\text{max}}$. Note that no structure in the data-set contained a separation greater then $10.0$. There are no maps for $r_{cut}=2.5$ (as with other $n_{\text{max}}$ and $l_{\text{max}}$ considered), since this cut-off is too low relative to the number of radial basis functions, preventing normalisation of the latter. Whilst the exact structure of the map varies with each set of parameters, there is a broad range (inside the bold, dashed, red perimeter) for which the key features are consistent with the maps produced using sorted lists of distances. Within this range of parameters, the large cluster of low-symmetry structures that is present in Figs.~\ref{fig:LJ38_labelled_surface} and \ref{fig:LJ38_3D_perspectives} is clear, the largest funnel, (d) in Fig.~\ref{fig:LJ38_labelled_surface}, is clearly revealed, and at least two other funnels are also discernible. For $\sigma$ values too large, the definition of the map diminishes, and the funnels are absorbed into the main cluster. For $\sigma$ values too small, the smooth variation in energy across the cluster of low-symmetry structures is lost.

Figs.~\ref{fig:LJ38_soap_params} (B) and (C) depict results using fewer basis functions: $n_{\text{max}}, l_{\text{max}} = 12, 6$ and $8, 4$, respectively. Comparing these to Fig.~\ref{fig:LJ38_soap_params} (A), we see that the band of suitable $\sigma$ shifts to lower values with lower $n_{\text{max}}$ and $l_{\text{max}}$. We also observe that the scope of acceptable parameters is of similar size for Figs.~\ref{fig:LJ38_soap_params} (A) and (B), but is significantly restricted in (C), for which only the mid-range $r_{cut}$ is able to produce the expected map features for any $\sigma$.

In general, the funnels labelled (a) and (b) in Fig.~\ref{fig:LJ38_labelled_surface} are not as well separated in the maps of Fig.~\ref{fig:LJ38_soap_params}. We attribute this to the smaller sample size used for these projections - 2D and 3D maps produced using SOAP from the full data-set, Fig.~\ref{fig:LJ38_soap_big} (which use $r_{cut} = 10.0, ~n_{\text{max}} = 15, ~l_{\text{max}} = 9, ~\sigma = 0.3$), provide clear separation of these funnels.

We conclude that, with the right choice of SOAP parameters, the picture of the $\text{LJ}_{38}$ energy landscape obtained through SHEAP is qualitatively equivalent for the two descriptors. Furthermore, we find that the range of safe/suitable SOAP parameters is relatively broad, and easy to identify for this system, via a scan of possible values, visualised through SHEAP.

\subsection{$\text{LJ}_{55}$}\label{LJ55}

\begin{figure*}[t]
    \centering
    \includegraphics[width=0.9\linewidth]{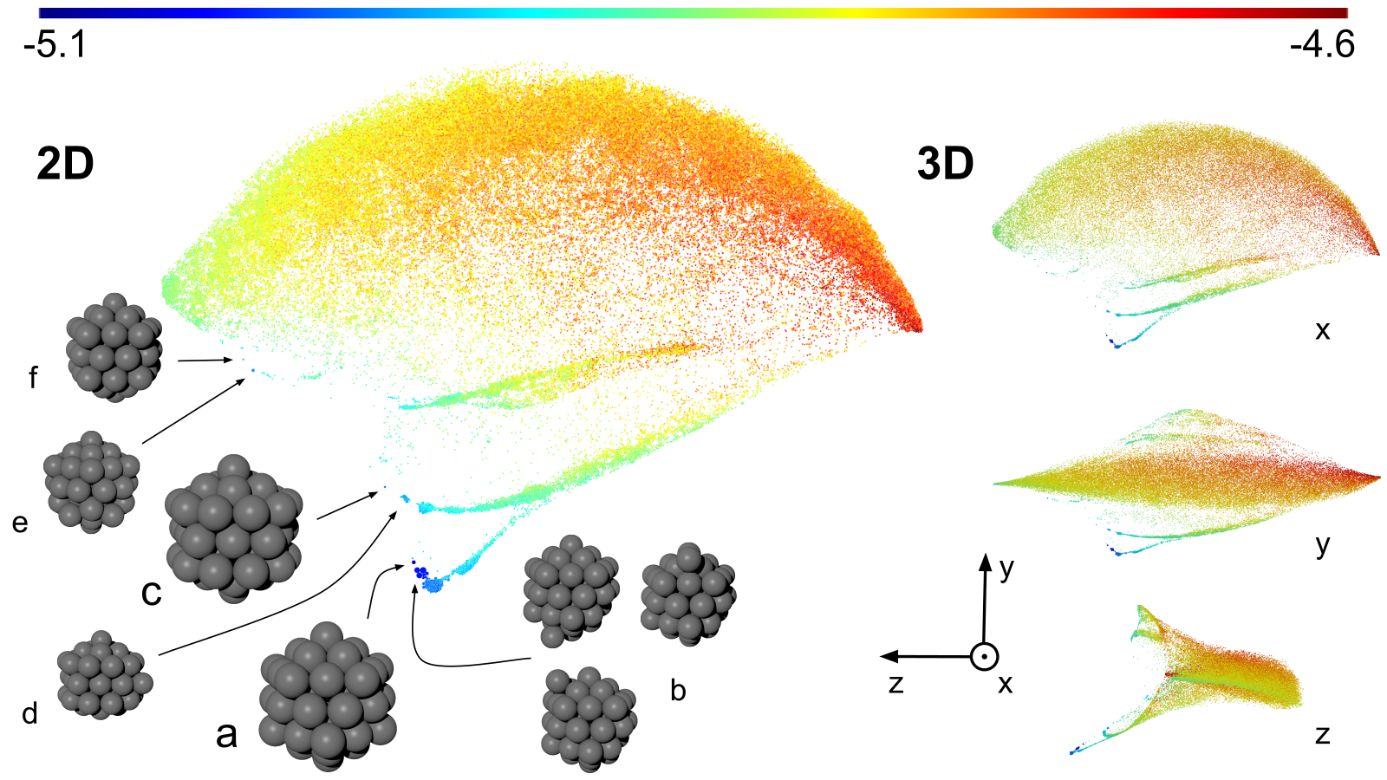}
    \caption{2D and 3D SHEAP maps for $49{,}813$ minima on the $\text{LJ}_{55}$ PES, generated with RSS, described by sorted lists of all pairwise distances. Projections were made using a perplexity of 30, with a minimum sphere radius $0.01$. The circles representing each structure are coloured according to energy per atom in Lennard-Jones units; see colour bar at top of figure. Basin volumes are represented by the area of each circle. (a)-(f) label selected structures in the landscape: (a) $\text{I}_\text{h}$, global minimum; (b) $\text{C}_\text{s}$, all have one surface atom relocated relative to global minimum, (c) $\text{D}_\text{5h}$, lowest energy structure in funnel branching off from that to global minimum; (d) $\text{C}_{1}$, one atom relocated relative to $\text{D}_\text{5h}$; (e) $\text{C}_\text{5v}$ and (f) $\text{C}_{5}$, high symmetry structures outside the main (sub-)funnels to $\text{I}_\text{h}$ and $\text{D}_\text{5h}$. Each viewpoint of 3D map looks down one of the axes displayed, as labelled; the axes have no other physical meaning.}
    \label{fig:LJ55}
\end{figure*}

$\text{LJ}_{55}$ is referred to as a “structure seeker”, meaning it has a very distinct global minimum that is easily located in a structure search, with no competing morphologies separated by high barriers \cite{sicher2011efficient}. Here we present a representation of this landscape obtained with SHEAP, highlighting its contrast to that of $\text{LJ}_{38}$.

Fig.~\ref{fig:LJ55} shows 2D and 3D SHEAP maps produced from minima on the $\text{LJ}_{55}$ PES. The search consisted of $50{,}000$ samples and used the same parameters and relaxation scheme as the above LJ systems, giving $49{,}813$ distinct minima.

There is clearly a funnelled structure to the arrangement of minima, terminating with the icosahedron and closely related configurations, such as those labelled (b), which all have the structure of the $\text{I}_\text{h}$ groundstate but with one external atom moved to a different surface site. However, the maps also suggest that describing the landscape as a single funnel would be an oversimplification. There is a second strand of structures which branches off from the funnel to the global minimum, leading to a relatively low-energy $\text{D}_\text{5h}$ configuration. This structure is related to the $\text{I}_\text{h}$ configuration via a twist about the 5-fold axis \cite{wales2003energy}. It stands out in the corresponding disconnectivity graph due to its relatively low energy and high barrier for transition to the global minimum \cite{doye1999evolution}; this picture is complemented by the detail in the structural similarity and sub-funnel arrangement provided by SHEAP.

Another band of (sub-)funnels is present, well separated from those leading to the $\text{I}_\text{h}$ and $\text{D}_\text{5h}$ structures in the 3D map (in a similar manner to the separation of (d) from (a), (b) and (c) in Fig.~\ref{fig:LJ38_labelled_surface}). However, these are all of relatively high energy (unlike (d) in Fig.~\ref{fig:LJ38_labelled_surface}), and so are not competetive with the global minimum.

Despite some similarities, it is apparent from the SHEAP maps why $\text{LJ}_{55}$ is a more straightforward problem than $\text{LJ}_{38}$ in terms of global optimisation. With $\text{LJ}_{55}$, the competing funnel(s)/strand(s) are shallow compared to that of the global minimum, and there are no similarly low-energy structures outside of the main funnel -  the $\text{D}_\text{5h}$ structure exceeds the global minimum by $\sim 0.1\epsilon$ per atom. With $\text{LJ}_{38}$, competing funnels are deep and terminate with structures similar in energy to the global minimum (only $\sim 0.01\epsilon$ per atom higher). Furthermore, for $\text{LJ}_{38}$ the most voluminous funnel is not the one containing the global minimum, and the maps also possesses a region of low-energy structures within the amorphous bulk of configurations, which is not the case for $\text{LJ}_{55}$.

\subsection{Carbon}\label{carbon}

\begin{figure*}[t]
    \centering
    \includegraphics[width=0.9\linewidth]{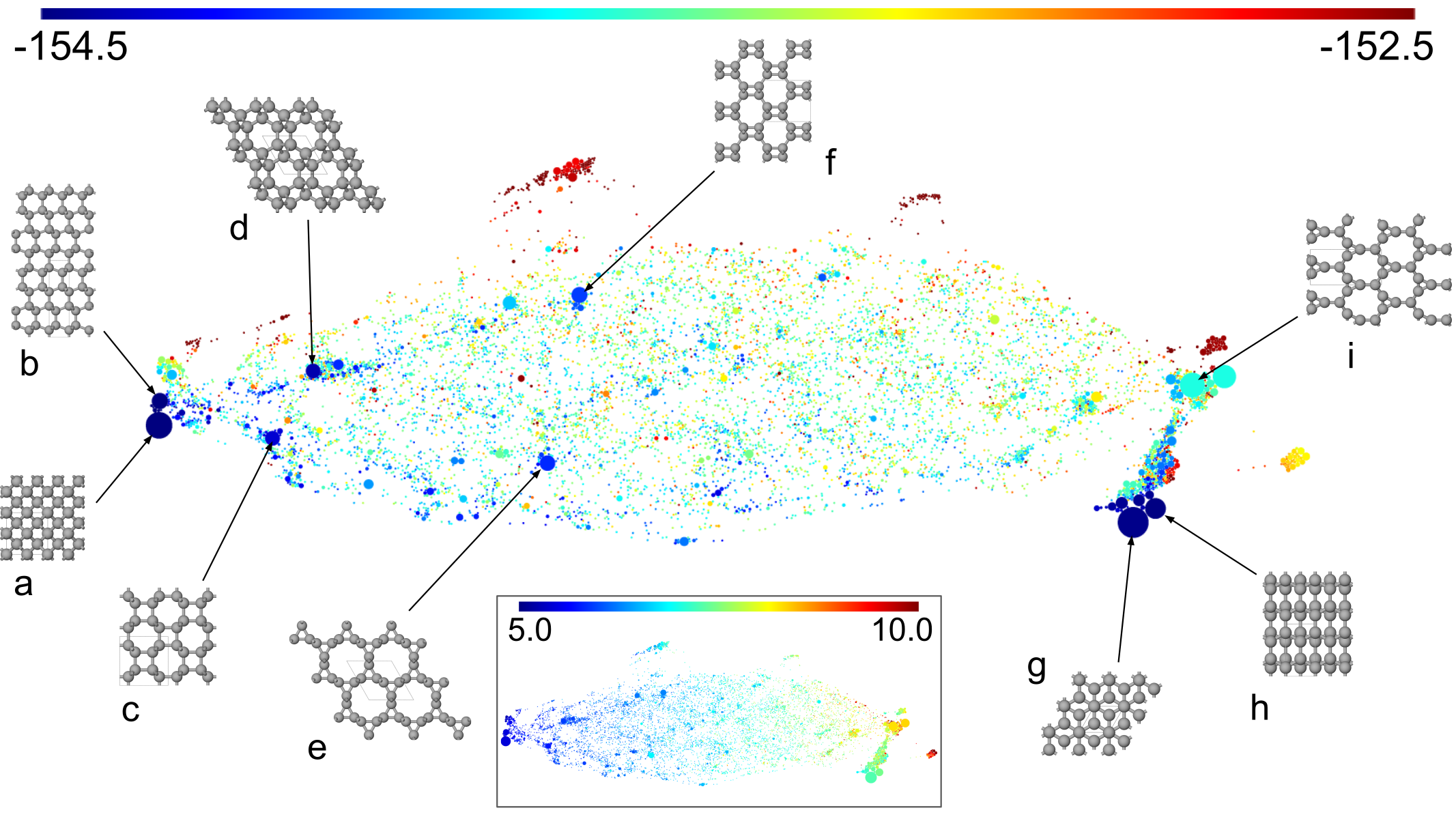}
    \caption{2D SHEAP map for $11{,}370$ structures of solid-state carbon, obtained with AIRSS at 10 GPa for a range of cell sizes - data-set taken from Materials Cloud Archive \cite{pickard2020airss}. Structures in the source-data represented as concatenated global SOAP vectors using universal ``smart'' parameters: $r_{\text{cut}},n_{\text{max}},l_{\text{max}},\sigma = \{2.4,8,4,0.3\}, ~\{2.0,8,4,0.25\}$. The projection was made using a perplexity of 30, with $R_0 = 0.0147$, as computed by Eqn.~\ref{R_0}. The circles representing each structure are coloured according to their energy per atom in eV; see colour bar at top of figure. Basin volumes are represented by the area of each circle. (a)-(n) label selected structures, including: (a) diamond, (b) lonsdaleite, (d) the chiral framework structure of \cite{pickard2010hypothetical}, (g) graphite, (h) graphene layers with non-graphite packing. Inset shows the same map, coloured according to volume per atom in $\text{\AA}^3$; see colour bar at top of inset.}
    \label{fig:carbon_labelled}
\end{figure*}

To demonstrate the applicability of SHEAP to realistic periodic systems, we consider a data-set of solid-state carbon structures, taken from the Materials Cloud Archive \cite{pickard2020airss}. The set contains $11{,}370$ structures obtained with AIRSS performed at 10 GPa for a range of cell sizes (up to 24 atoms), with 2 to 4 symmetry operations imposed on the pre-relaxed structures. All structures are based on sp/sp2/sp3 hybridised carbon, but they exhibit a wide range of topologies. The interactions in this system contain significant 3-body contributions \cite{deringer2017machine}, meaning the sorted list of pairwise distances is not a suitable structure descriptor. Hence, we use SOAP, which is capable of describing these relationships completely \cite{bartok2013representing}.

As demonstrated with $\text{LJ}_{38}$, the SOAP descriptor has a number of parameters which effect the appearance of a SHEAP map. With that model system, we were able to  obtain suitable values relatively easily via a scan of the parameter space. However, it would of course be preferable to have a systematic way to obtain good parameter choices for any given system. For systems of real atoms, we find a satisfying solution is the heuristic approach of Cheng et al. \cite{cheng2020mapping}, providing a set of so called ``universal'' parameters. These come in 3 flavours; ``smart'', which computes a suitable set of parameters by considering the typical and minimum bond lengths of each of the species involved, ``minimal'', which is similar to smart, but uses fewer basis functions, and ``longrange'', which again is similar to smart, but aims to capture more of the long-range structure. For the ``smart'' and ``longrange'' cases, two sets of SOAP parameters are computed, at slightly different length scales. A single structure descriptor is constructed by concatenating the global SOAP vectors for each set. Here, we provide a SHEAP map for the above carbon data set produced using the ``smart'' parameters - see Fig.~\ref{fig:carbon_labelled}.

Most structures lie within a single, elongated arrangement. Several have distinctly large basins; most of these are low-energy and lie towards the edge of the distribution. Many of these correspond to well known carbon arrangements, including diamond (a) and graphite (g), which reside at opposite ends of the map, as well as lonsdaleite (b), and the chiral framework structure (d) of Ref.~\cite{pickard2010hypothetical}. These and other standout clusters have been labelled to provide a representative picture of the distribution of structures.

Comparing the SHEAP map for carbon to those for the LJ systems highlights key differences in their energy landscapes. With the LJ systems, neighbouring minima tend to be close in energy. With carbon, there is a large degree of ``mixing'' between high- and low-energy structures; basins of high-energy minima are often surrounded by those of much lower energy, and vice versa. Furthermore, the variation in the energy of minima across the carbon map shows less clear long-range trends than for the LJ clusters. Whilst the LJ landscapes possess clear funnels of minima, such features do not appear to be prevalent in the carbon PES. Through this comparison, it is easy to see why LJ landscapes lend themselves very well to structure prediction algorithms such as simulated annealing \cite{kirkpatrick1983optimization} and basin-hopping \cite{li1987monte,wales1997global}, which utilise the presence of funnels to locate progressively lower energy minima, whilst the study of carbon has required more heuristic approaches, such as in Ref.~\cite{shi2018stochastic}.

As illustrated in the inset of Fig.~\ref{fig:carbon_labelled}, in contrast to the energy variation, the volume per atom of each carbon structure follows a clear global trend across the map. This demonstrates the additional complexity in the energy landscape resulting from 3-body interactions, leading to less correlation between the energy and density than is seen for a simple pairwise interaction.

\subsection{C+H+N+O}\label{CHNO}

\begin{figure*}[t]
    \centering
    \includegraphics[width=0.9\linewidth]{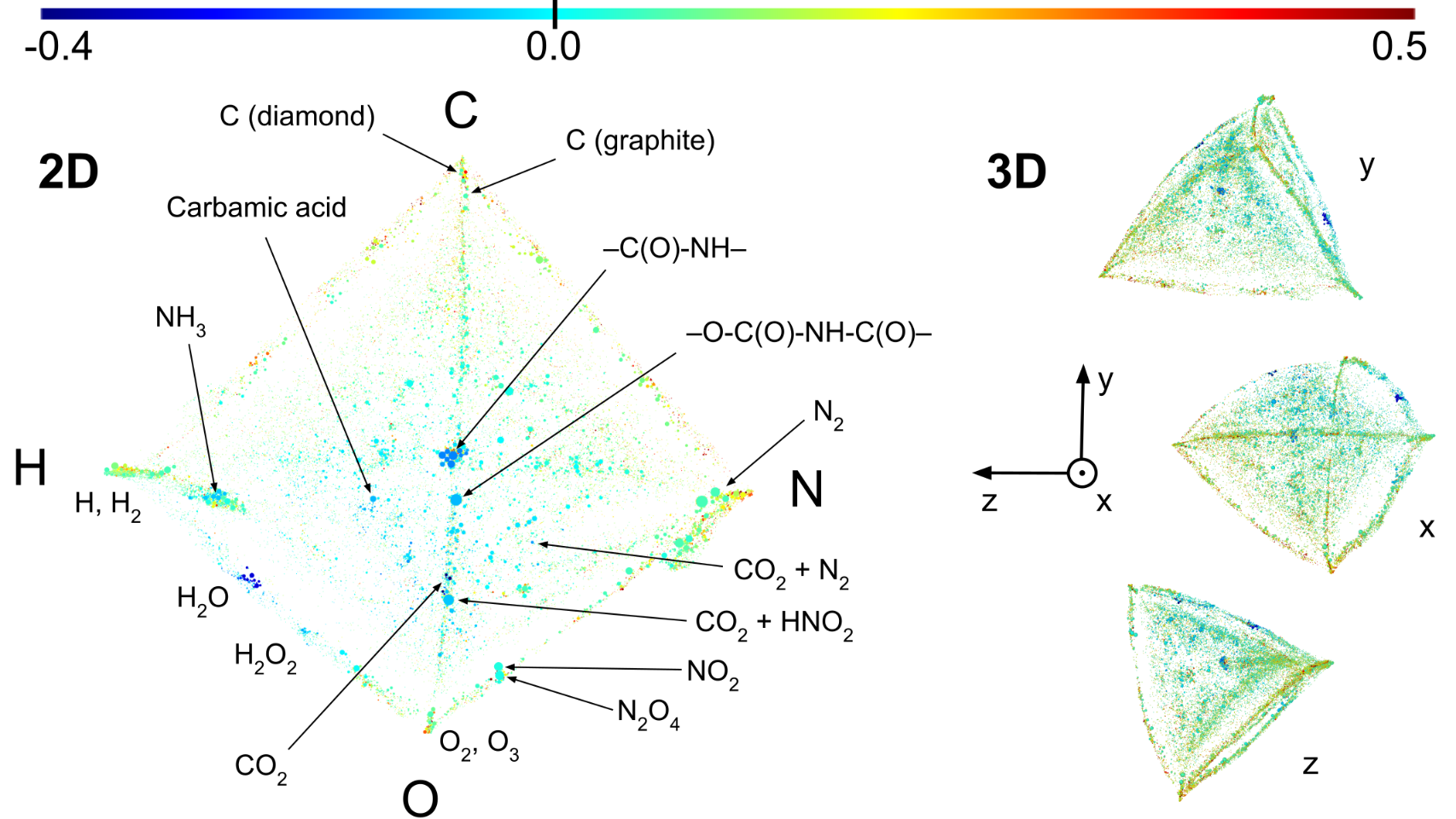}
    \caption{2D and 3D SHEAP maps for $21{,}239$ structures containing carbon, hydrogen, nitrogen and oxygen at varying stoichiometries, obtained with AIRSS performed at 1 GPa, taken from the Materials Cloud Archive \cite{pickard2020airss}. Structures in the source-data represented as concatenated global SOAP vectors using ``smart'' universal parameters: $r_{\text{cut}},n_{\text{max}},l_{\text{max}},\sigma = \{2.7,8,4,0.33\}, ~\{2.0,8,4,0.25\}$. Projections were made using a perplexity of 30, with $R_0$ computed by Eqn.~\ref{R_0}. Circles/spheres representing each structure are coloured according to their formation energy per atom in eV; see colour bar at top of figure. Basin volumes are represented by the area/volume of each sphere/circle. Each viewpoint of the 3D map looks down one of the axes displayed, as labelled; the axes have no other physical meaning. Clusters/regions containing polymorphs of various molecular species are labelled in the 2D map.}
    \label{fig:CHNO_labelled}
\end{figure*}

As a final demonstration of SHEAP, we present results obtained for a data-set of structures containing multiple species. The set considered consists of $50{,}000$ solid-state structures containing carbon, hydrogen, nitrogen and oxygen (the four most common elements in organic chemistry and biology) at varying stoichiometries, which reduced to $21{,}239$ distinct structures following a similarity check. These configurations, taken from a larger set on the Materials Cloud Archive \cite{pickard2020airss}, were generated with AIRSS performed at 1 GPa. Structures in the source data-set are represented as concatenated global SOAP vectors, using the ``smart'' universal parameters, including the crossover terms between different species. Presented in Fig.~\ref{fig:CHNO_labelled} are 2D and 3D SHEAP maps for this system.

The 2D map contains four distinct corners, each corresponding to arrangements of one of the four elements, as labelled. One can easily identify clusters of basins corresponding to low-energy configurations of binary molecules, most notably $\text{H}_2\text{O}$, $\text{CO}_2$, and $\text{NH}_3$. Intuitively, each of these molecular structures resides closest to the corner corresponding to the element it contains with the highest stoichiometry. The edges of the map represent structures containing predominantly the elements of the adjacent corners, also in agreement with intuition. The same is true for the structures occupying the regions between opposite corners (H, N and C, O). The lowest energy structures containing each element in equal stoichiometry correspond to polymers of isocyanic acid (O=C=NH), which reside close to the centre of the map, as labelled. A few other standout clusters/basins have been labelled, according to the molecule(s) whose polymorphs are are mapped there, to provide a representative picture of the distribution of structures and stoichiometries.

The 3D map also contains a distinct corner for each of the 4 elements. However, the extra dimension facilitates the emergence of an edge between every pair, and provides clearer separation of different stoichiometries, particularly for structures residing towards the centre of the 2D map. The result resembles an irregular tetrahedron, with slight outwards curvature of the faces and edges. This is reminiscent of a quaternary convex hull plot \cite{flores2020perspective}; it is striking that the similarities to this construction have emerged naturally from our projection of the energy landscape.

Both maps depict a higher density of structures at the edges (and between opposite corners in 2D), where the binary (and unary) configurations reside. This demonstrates that SHEAP is able to highlight any non-uniformity or biasing in the sampling of a configuration space.

\subsection{Cost vs.~dimensionality}\label{cost_dimensionality}

\begin{figure}[t]
    \centering
    \includegraphics[width=\linewidth]{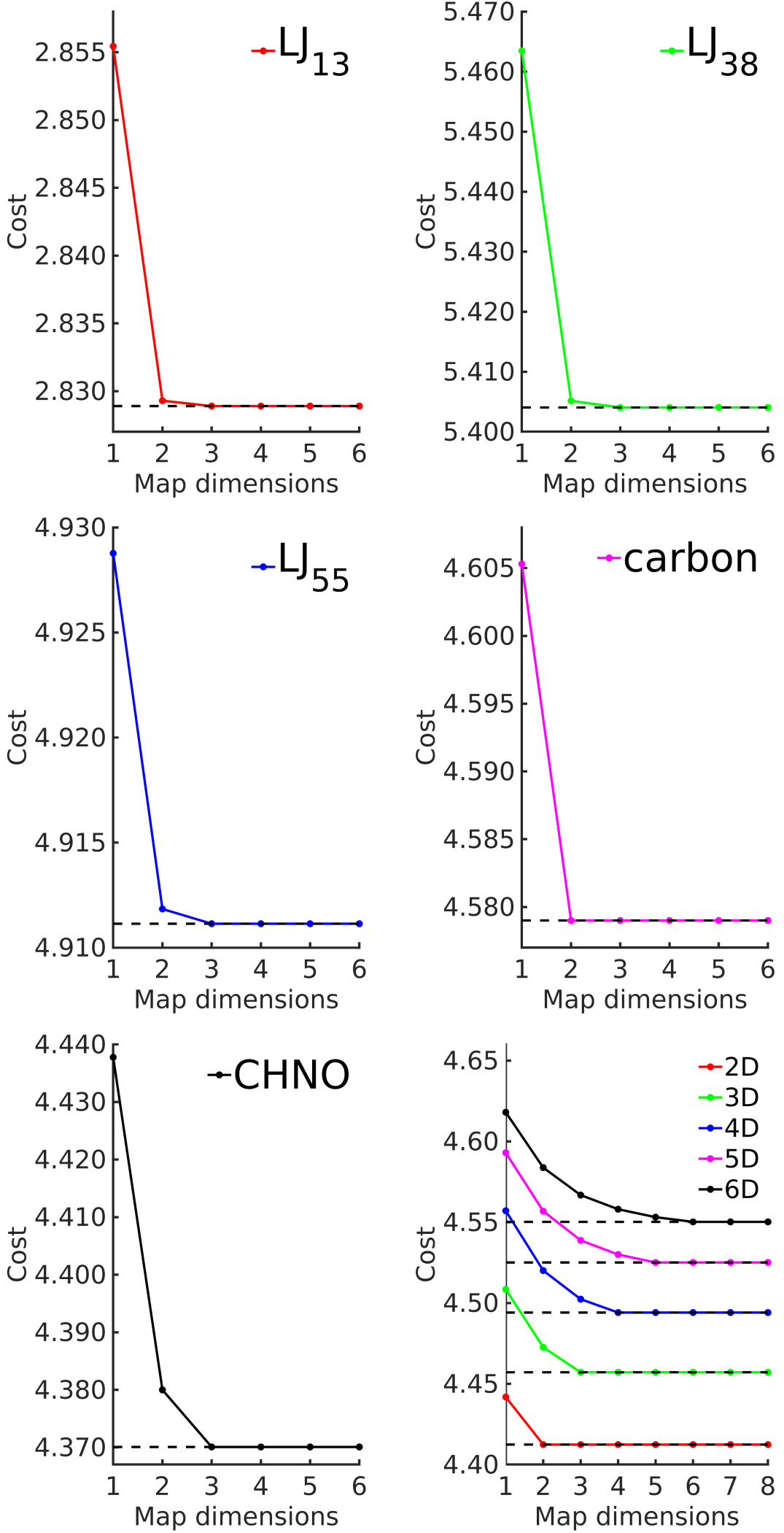}
    \caption{Lowest cost obtained through SHEAP projections (without hard-sphere interactions) of data-sets of $10{,}000$ local minima for five systems, plotted against the number of map dimensions. Also shown are (lowest) cost vs.~map dimension plots for $10{,}000$ points drawn uniformly at random inside the unit hypercubes of 2-6 dimensions, on a single set of axes. Each SHEAP projection used a perplexity of 30.}
    \label{fig:cost_dimensionality}
\end{figure}

In Sections \ref{LJ38}, \ref{LJ55} and \ref{CHNO} we demonstrated that the visualisation quality produced by SHEAP can be improved by projecting into three dimensions, rather than two. How do we know that there are no important features in a given minima distribution that would be revealed if we were able to view a projection into higher dimensions? We address this by considering the optimal cost (in the absence of the hard sphere interaction) achieved by projections into a range of map dimensionalities.

This analysis was conducted for data-sets of $10{,}000$ local minima of each of the systems studied above, as well as data-sets consisting of $10{,}000$ points drawn uniformly at random inside the unit hypercubes of 2 to 6 dimensions. For the structural data, 10 independent SHEAP projections were carried out, without switching on the hard sphere interaction, for 1 to 6 map dimensions, $s$. For the points inside hypercubes, SHEAP projections were conducted for up to 8 dimensions. The lowest cost obtained at each map dimension, for each data-set, is plotted against $s$ in Fig.~\ref{fig:cost_dimensionality}.

Note that the cost is restricted from approaching zero as $s \longrightarrow D$ (and beyond) by the use of different functional forms for the weights in the high and low dimensions; there is an inherent cost associated with the use of different distributions, even when $s \geq D$.

For all data-sets of structural minima, we observe a dramatic reduction in the (lowest) relative cost in going from 1D to 2D. We also see a non-negligible reduction from increasing the map dimensionality from two to three (except in the case of carbon), which is most pronounced for C+H+N+O. However, it appears that in each case there is nothing of significance to be gained in going beyond 3D.

We compare to the plots for uniformly drawn random points within hypercubes of varying dimensionality. According to SHEAP, in order to faithfully represent the distribution of points in the hypercube of dimensionality $D$ (which must contain long-range $D$-dimensional features), at least $s = D$ map dimensions are required. For each cube, using fewer than $D$ dimensions in the map leads to a penalty in the optimal cost that is obtained, which increases with decreasing map dimensionality.

These results suggest that the distribution of minima on the landscapes we have considered are inherently low dimensional (according to the given combination of structure descriptor and distance metric), providing validation for our approach to producing visual representations of high-dimensional PESs.

\section{Discussion}

The striking observation of local minima being approximately distributed across low-dimensional manifolds within configuration space could have profound consequences for structure prediction. The ability to sample directly from this manifold would be a highly efficient means of searching for stable structures. Unfortunately, SHEAP (as with all algorithms in the same family) does not yield a transformation matrix between the original and reduced spaces, and the latter is not defined with respect to a meaningful set of axes. Thus, there is no clear way to generate new structures on the manifold learned by SHEAP. However, our findings are promising for the use of other machine learning methods such variational auto-encoders \cite{kingma2013auto} (VAEs) and generative adversarial networks \cite{goodfellow2014generative} (GANs) in structure prediction for generating candidate structures (see Ref.~\cite{kim2020generative}), and suggest that learning from previously located minima would be an effective strategy. This could be achieved by pairing with a non-learning based approach such as AIRSS. SHEAP could be deployed to illuminate differences in the sampling of a given surface by different approaches, or by different parameter choices within a given approach.

Other pioneering work in developing a greater understanding of energy landscapes through visualisation is Ref.~\cite{oganov2009quantify}, which also analyses similarities between local minima to reveal topological features. Oganov and Valle employ a fingerprint structure descriptor based on pairwise separations of atoms \cite{valle2008crystal}, and compute the similarities between fingerprints with a cosine distance metric. They use this similarity measure to construct plots of energy vs. distance to the global minimum for local minima on a given landscape, as well as mappings of the minima distributions which aim to produce ``distances between points in the graph that are maximally close to the distances between the corresponding fingerprints'' (see multi-dimensional scaling \cite{torgerson1952multidimensional}). Their analysis reveals funnels in the landscapes of crystalline systems of few atoms per unit cell. Whilst Ref.~\cite{oganov2009quantify} does not focus on the quality of the mapping between their description of a given PES and the low-dimensional space it is visualised in, and only deals with relatively small sample sizes (a few thousand structures), the tools developed there have been demonstrated to facilitate improved structure searching \cite{lyakhov2010predict}. Dimensionality reduction by SHEAP furthers the potential of such an approach.

In this work we have represented real crystalline systems using SOAP. However, it is important to note that the development of good structure descriptors is an active field, and there are numerous justifiable choices we could have made for the descriptor-metric pairing. Of course, the entire form of a map is dependent on the choice of descriptor and distance metric, and a map's features cannot be interpreted without knowledge of the descriptor-metric pair used. With SOAP, a global reprentation of a given structure is constructed by averaging the local fingerprints of each atomic environment. As with any instance of averaging, this step introduces a loss of structural information, which becomes severe for large unit cells. A potential solution is to build on the work of Ref.~\cite{zhu2016fingerprint}. Rather than comparing averaged SOAP descriptors via a Euclidean distance metric, one could use the Hungarian algorithm to best match the atomic environments in a pair of structures, summing over the individual distances between matching environments. Any such, more sophistocated approach would need to be assessed with regard to the balance between increased fidelity of the map versus the computational cost of the algorithm.

\section{Conclusion}\label{conclusions}

In conclusion, SHEAP enables us to produce meaningful and interpretable low-dimensional representations of a range of PESs, requiring only the structural information contained in minima data obtained from random structure searches as input, providing fresh insight into their topologies. The maps presented contain distributions and clusters of structures that we have been able to rationalise, and reveal long range topological features, including funnels (or the lack thereof).

By evaluating the costs at higher map dimensionalities, we have assessed the validity of SHEAP projections into the low number of dimensions that we can comprehend. In doing so, we have revealed an intrinsic low-dimensionality to the distributions of minima, as represented by the descriptors/metrics considered, across both model PESs, and those corresponding to realistic, solid-state systems. We observe that with all systems studied, no more than three map dimensions are required to represent all of the structural variation captured in our descriptions of the local minima, with two map dimensions also enabling useful projections.

We have demonstrated that SHEAP has the potential to be a useful tool in the field of structure prediction (and beyond), providing a convenient and concise way to visually represent and compare the large quantities of data that are generated in structure searches. The insight gained from these projections could provide inspiration for approaches to improving/optimising sampling of a given landscape for a desired purpose.

\begin{acknowledgments}

We thank Daan Frenkel, Joseph Nelson, and Bonan Zhu for useful discussions. BS acknowledges the EPSRC CDT in Computational Methods for Materials Science for funding under grant number EP/L015552/1. CJP has been supported by the Royal Society through a Royal Society Wolfson Research Merit award. This work was performed using resources provided by the Cambridge Service for Data Driven Discovery (CSD3), operated by the University of Cambridge Research Computing Service (www.csd3.cam.ac.uk).

\end{acknowledgments}

\begin{appendix}
\section{UMAP vs. t-SNE}\label{UMAP_tsne}

For algorithms that follow the framework outlined in Section \ref{framework}, the optimisation procedure is usually dictated by a cost-function gradient that can be separated into a sum of attractive and repulsive pairwise interactions between map points;
\begin{equation}
    \frac{d C}{d Y} = \sum_{i} \sum_{j > i} F_{\text{att.}}(\mathbf{y_i}, \mathbf{y_j}) - F_{\text{rep.}}(\mathbf{y_i}, \mathbf{y_j}) \; .
\end{equation}
The t-SNE algorithm exploits this with its use of early exaggeration to accelerate the optimisation \cite{maaten2008visualizing}; this technique works by the temporary enhancement of the attractive contribution to the gradient, achieved through multiplying these terms by some positive scalar $\alpha > 1$ for the first few hundred iterations of optimisation, promoting the formation of clusters in the projected data.

In t-SNE, the cost function measuring the faithfulness with which the probabilities $q_{ij}$ model their corresponding $p_{ij}$ is the Kullback-Leibler (KL) divergence of the former distribution from the latter \cite{maaten2008visualizing}:
\begin{equation}
    C_{\text{t-SNE}} = \sum_{i,j} p_{ij} \log \left( \frac{p_{ij}}{q_{ij}} \right) \, .
    \label{KL_tsne}
\end{equation}
The attractive contribution to the gradient in t-SNE is a direct result of the form of the cost function, and is what drives the algorithm to accurately reproduce the local connectivity of the source data in the projected map. The repulsive contribution is a result of the constraint that the low-dimensional weights $q_{ij}$ sum to 1. This can be incorporated either through the treatment of the optimisation with an additional Lagrangian multiplier term, or directly in the definition of the weights. Without this constraint, the KL-divergence would be minimised by all points in the map collapsing onto one another.

UMAP chooses to minimise the fuzzy set cross-entropy between the two distributions \cite{mcinnes2018umap1}:
\begin{equation}
    C_{\text{UMAP}} = \sum_{i,j} p_{ij} \log \left( \frac{p_{ij}}{q_{ij}} \right) + (1-p_{ij}) \log \left( \frac{1-p_{ij}}{1-q_{ij}} \right) \; .
    \label{ce_UMAP}
\end{equation}
The first term is identical to the KL-divergence used by t-SNE, and produces the attractive contribution to the gradient. In UMAP, this acts only between map points that are considered to be nearest neighbours in the source data, meaning $p_{ij} = 0$ for all pairs of points $i$ and $j$ for which neither is considered one of the $k^{th}$ nearest neighbours ($k$ is a user-specified hyperparameter) of the other, according to the defined distance metric. The second, additional term introduces a repulsion between every pair of map points that is greater the further apart the corresponding points are in the source data. This choice of cost function is the key reason that UMAP is able to provide improved performance over t-SNE.

Firstly, in t-SNE the repulsive term in the gradient appears only because of the constraint that the low-dimensional weights $q_{ij}$ sum to one, which requires the re-normalisation of all of these weights at each iteration of the optimisation procedure. In UMAP, the cost function introduces a repulsive contribution to the gradient without the need for this constraint, meaning the costly normalisation step can be avoided.

Secondly, the repulsive term contribution to the gradient in UMAP is more dependent on the actual data than that in t-SNE. Thus, UMAP tends to do a better job than t-SNE in producing a faithful representation of the global structure of the source data.

Finally, the removal of the need to normalise the weights $q_{ij}$ means that the cost function is amenable to optimisation via SGD, which replaces the true gradient with an estimate that requires only a stochastically chosen subset of the components to be evaluated. UMAP implements an efficient approximate SGD algorithm which makes use of probabilistic edge sampling and negative sampling \cite{mcinnes2018umap1}. The improvements in speed and scalling achieved by UMAP over t-SNE are largely a result of the use of SGD, rather than some other full gradient descent method.

However, a fairly subtle caveat arises with UMAP's implementation of SGD. The repulsive part of the interaction between map points is treated with negative sampling \cite{mikolov2013distributed}. Whilst this vastly reduces the number of repulsive interactions that need to be evaluated, it also has the effect of de-exaggerating the repulsion between the map points, relative to their interaction as described by the true cost function. We find that this is crucial to UMAP producing a meaningful projection. Because of this, UMAP's optimisation scheme does not actually minimise the defined cost function, and the algorithm only ``converges'' because the step size is forced to go to zero. Thus, whilst UMAP is very efficient, and produces low-dimensional projections that appear to improve on those by the t-SNE algorithm, it is rather unsatisfactory in that the cost function defined within the algorithm is not truly minimised in the construction of the final map.

\section{Choosing $\mathbf{R_0}$}\label{R_0}

The choice of the minimum sphere/circle radius $R_0$ is critical. If, in the absence of the hard spheres, the separation between a given pair of map points is smaller than the sum of their hard sphere radii, then the introduction of the hard sphere interaction will force these structures apart. Thus, if $R_0$ is large enough for this to be the case for a single pair of structures, the introduction of hard spheres results in a distortion of the layout of the map. On the other hand, if $R_0$ is too small relative to the spread of data in the map, some (or all) of the spheres may be barely visible. Hence, a good choice for $R_0$ compromises between ensuring that all of the structures in the map are clearly visible, whilst minimising the distortion of the layout of structures.

By default, the value of $R_0$ is computed in the same iteration at which the hard sphere interaction is switched on, as a fraction of the average separation of the map points from their centre of mass:
\begin{equation}
    R_0 = \frac{2^s}{10} \left( \frac{N}{\sum_{i=1}^N c_i} \right)^{1/s} \langle ||\mathbf{y}_i - \mathbf{y}_{\text{CoM}} || \rangle_{\text{HS}} \; ,
    \label{R_0}
\end{equation}
where
\begin{equation}
    \langle ||\mathbf{y}_i - \mathbf{y}_{\text{CoM}} || \rangle_{\text{HS}} = \frac{1}{N} \sum_{i=1}^N ||\mathbf{y}_i - \mathbf{y}_{\text{CoM}} ||_{\text{HS}} \; ,
\end{equation}\\
and the pre-multiplying factor was selected by empirical investigation, and may be altered in future. Alternatively, the value of $R_0$ can be input by the user, allowing interactive feedback.

\section{Random projection}\label{random_projection}

\begin{figure*}[t]
    \centering
    \includegraphics[width=0.9\linewidth]{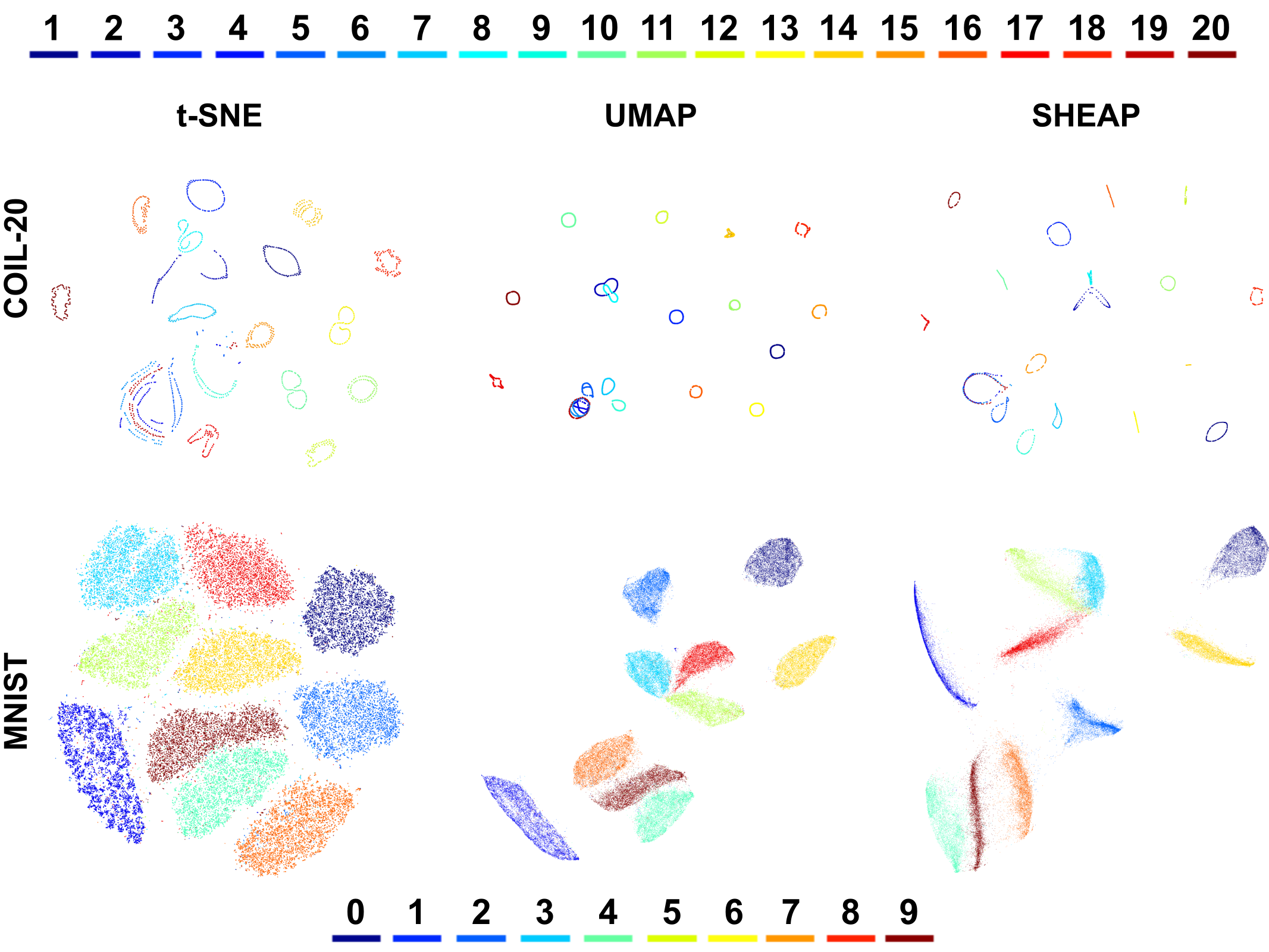}
    \caption{Projections of the COIL-20 \cite{nene1996columbia} and MNIST \cite{lecun1998gradient} data-sets made by our implementations of the t-SNE, UMAP, and SHEAP algorithms. The legend at the top of the figure labels distinct objects in the COIL-20 data-set; the legend at the bottom labels distinct digits in the MNIST data-set.}
    \label{fig:standard_datasets}
\end{figure*}

Random projection relies on the Johnson-Lindenstrauss lemma \cite{johnson1984extensions}, which states that any set of $N$ points in a high-dimensional Euclidean space can be embedded into a $k \geq \mathcal{O} \log (N / \epsilon^2)$ dimensional Euclidean subspace without distorting the distances between any pair of points by more than a factor of $1 \pm \epsilon$, for any $0 < \epsilon < 1$. It can be proven that if the $k$-dimensional subspace is constructed by the generation of $k$ random but mutually orthogonal and normalised vectors in the high-dimensional space, then the this will be achieved with high probability. Furthermore, it is found that in practice, even if embedding into a low-dimensional space of dimension $s << \mathcal{O} \log (N / \epsilon^2)$, one can often still obtain something useful from a random projection. In our implementation of random projection, a source data-point $\mathbf{x}$ existing in the original $D$-dimensional space is projected onto an $s$-dimensional subspace by transformation via a $s \times D$ matrix $R$ whose $s$ rows are orthonormal vectors in the original space, to produce the projected data-point:
\begin{equation}
    \mathbf{y}^{RP} = R \mathbf{x} \; .
\end{equation}
In addition, because our descriptor vectors contain only positive entries (they are pairwise distances), we know that all of the data-points exist in the positive orthant of the $D$-dimensional hyperspace, meaning only a $1/2^{d}$ fraction of the total space is accessible. Hence, we actually choose the row vectors comprising $R$ to contain only positive entries.

\section{Qualitative comparison of SHEAP to t-SNE and UMAP}\label{standard_datasets}

\begin{figure}[t!]
   \centering
   \includegraphics[width=\linewidth]{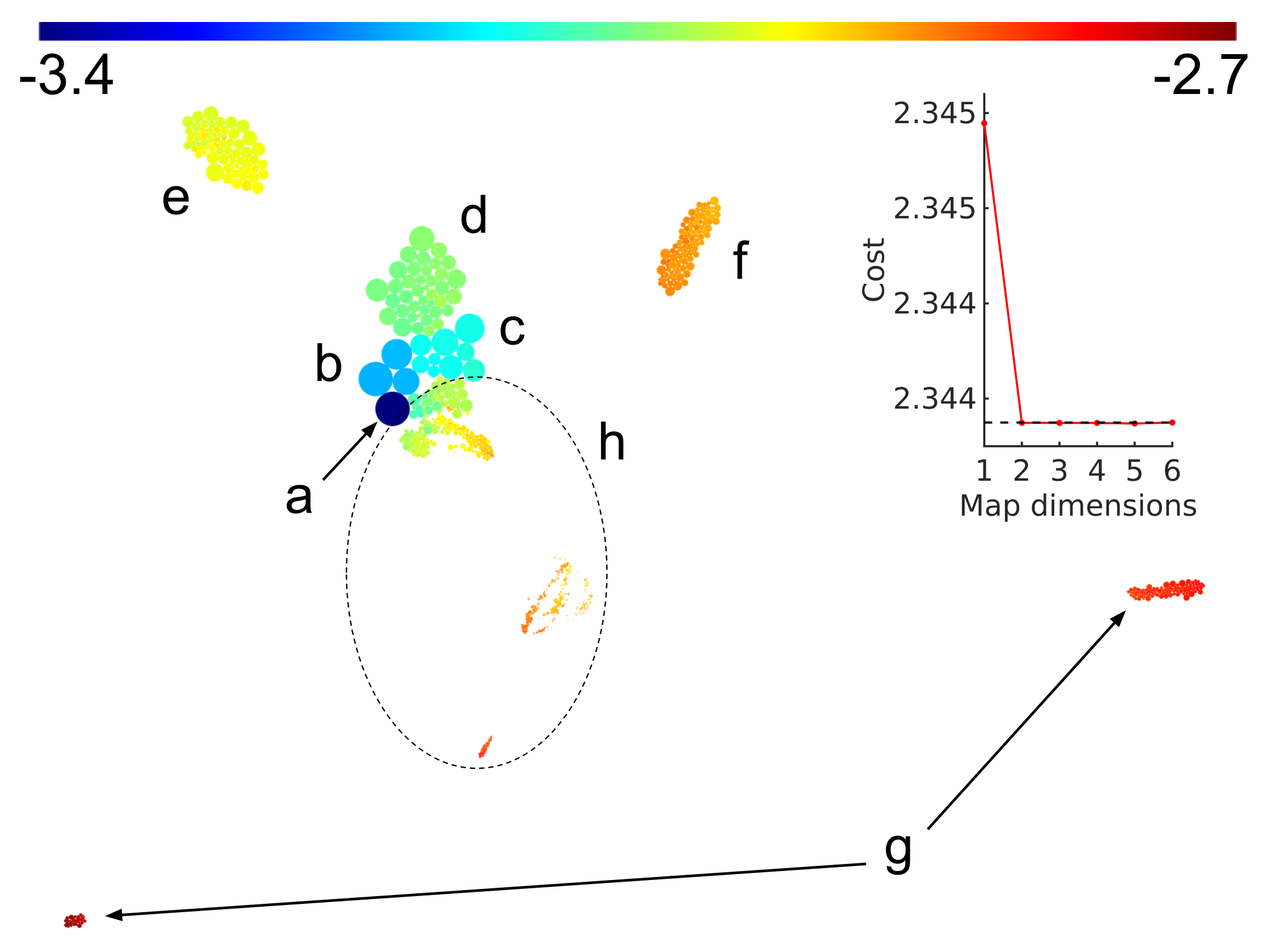}
   \caption{Labelled 2D SHEAP map produced from local minima on the $\text{LJ}_{13}$ PES (the same set as depicted in Fig.~\ref{fig:LJ13_labelled_surface}), described by sorted lists of inverse pairwise distances. The projection was made using a perplexity of 30, with a minimum sphere radius $R_0 = 0.01$. The circles representing each structure are coloured according to energy per atom in Lennard-Jones units; see colour bar at top of figure. Basin volumes are represented by the area of each circle. Groups of structures are labelled according to their clustering in Fig.~\ref{fig:LJ13_labelled_surface}; the dashed line encloses structures collectively labelled (h). Also shown is a plot of the lowest cost (without hard sphere interactions) obtained by projections of a subset of this data-set into a range of map dimensionalities.}
   \label{fig:LJ13_inv_r}
\end{figure}

The visualisation quality of SHEAP is compared against our implementations of t-SNE and UMAP on two standard image data-sets: COIL-20 \cite{nene1996columbia}, and MNIST \cite{lecun1998gradient}. In general, the improvements offered by UMAP over t-SNE are maintained with SHEAP.

The COIL-20 \cite{nene1996columbia} data-set contains $1,440$ images, produced by individually placing 20 objects on a motorised turn-table (against a black background) and photographing them at $5\degree$ intervals, up to a complete rotation. Each image is grey-scale, containing $128 \times 128$ pixels, and is treated as a $16{,}384$-dimensional vector. This data-set is expected to contain well defined clusters, corresponding to the sets of images of each object. Furthermore, the sub-manifold on which each cluster lies should have a relatively intuitive form - objects containing no symmetry about the axis of rotation are expected correspond to a closed loop of points, with consecutive images of a given object neighbouring one another. In Fig.~\ref{fig:standard_datasets}, both UMAP and SHEAP do a better job than t-SNE at forming the clusters corresponding to each object, and keep more of the loops intact. The biggest difference between the UMAP and SHEAP projections is that SHEAP produces larger/more frequent distortions of these closed loops - we argue that these distortions make intuitive sense, occurring when there exist symmetries in the objects being imaged.

The MNIST \cite{lecun1998gradient} (training) data-set contains $60{,}000$ grey-scale $28 \times 28$ images of handwritten digits (0 to 9), each treated as a $784$-dimensional vector. A given data-point is expected to be neighboured by other writings of the same digit, and the data-set should contain well-defined clusters corresponding to the distinct digits. These are expected to be arranged such that similar digits reside close together, and dissimilar ones are well separated (a handwritten 3 is often hard to distinguish from an 8, but never mistaken for a 4). In Fig.~\ref{fig:standard_datasets}, all three algorithms do a good job of clustering the images corresponding to each digit. However, UMAP and SHEAP capture more of the global relationships between the clusters, providing much clearer separations between them, and grouping together clusters of similar digits (3,5,8, and 4,7,9) in similar layouts. A noticeable difference between UMAP and SHEAP's projections is that the latter seems to contain thinner, denser bands of points in clusters that are particularly elongated. Nonetheless, the two plots are qualitatively very similar.

\section{$\text{LJ}_{13}$ described with inverse distances}\label{LJ13_inv}

Fig.~\ref{fig:LJ13_inv_r} shows a 2D SHEAP map for the same set of $\text{LJ}_{13}$ minima as considered in Section \ref{LJ13}. The difference here is that, rather than describing structures with sorted lists of all distances between pairs of atoms, we use sorted lists of all inverse distances. We observe similar clustering of structures as in Fig.~\ref{fig:LJ13_labelled_surface}; groups of structures in Fig.~\ref{fig:LJ13_inv_r} are labelled according to the corresponding cluster in Fig.~\ref{fig:LJ13_labelled_surface}. However, the map obtained from inverse distances does not result in a consistent global layout of the minima, failing to reproduce the expected single funnel distribution across all minima. We speculatively attribute this to the emphasis the descriptor places on the shortest distances in the structures, resulting in a poorer representation of global information.

Also in Fig.~\ref{fig:LJ13_inv_r} is a plot of the lowest cost (in the absence of hard sphere interactions) achieved with projections into a range of map dimensionalities, similar to the plots in Fig.~\ref{fig:cost_dimensionality}. Again, this analysis was conducted for a reduced data-set of $10{,}000$ minima, with 10 independent SHEAP projections carried out for each number of map dimensions. As with the other structural data-sets considered, no more than 3 map dimensions are required for convergence. However, here, 3 map dimensions does not produce a lower cost than 2, whereas a slight reduction was observed with the sorted list of distances descriptor for the same system. This is likely due to the less well represented global features of the landcape.

\end{appendix}

\end{document}